\PassOptionsToPackage{unicode}{hyperref}
\PassOptionsToPackage{hyphens}{url}
\PassOptionsToPackage{dvipsnames,svgnames,x11names}{xcolor}
\documentclass[
  9pt,
  a4paper,
]{article}

\usepackage{amsmath,amssymb}
\usepackage{lmodern}
\usepackage{iftex}
\ifPDFTeX
  \usepackage[T1]{fontenc}
  \usepackage[utf8]{inputenc}
  \usepackage{textcomp} 
\else 
  \usepackage{unicode-math}
  \defaultfontfeatures{Scale=MatchLowercase}
  \defaultfontfeatures[\rmfamily]{Ligatures=TeX,Scale=1}
\fi
\IfFileExists{upquote.sty}{\usepackage{upquote}}{}
\IfFileExists{microtype.sty}{
  \usepackage[]{microtype}
  \UseMicrotypeSet[protrusion]{basicmath} 
}{}
\makeatletter
\@ifundefined{KOMAClassName}{
  \IfFileExists{parskip.sty}{%
    \usepackage{parskip}
  }{
    \setlength{\parindent}{0pt}
    \setlength{\parskip}{6pt plus 2pt minus 1pt}}
}{
  \KOMAoptions{parskip=half}}
\makeatother
\usepackage{xcolor}
\usepackage[top=25.4mm,bottom=25.4mm,right=25.4mm,left=25.4mm]{geometry}
\ifx\paragraph\undefined\else
  \let\oldparagraph\paragraph
  \renewcommand{\paragraph}[1]{\oldparagraph{#1}\mbox{}}
\fi
\ifx\subparagraph\undefined\else
  \let\oldsubparagraph\subparagraph
  \renewcommand{\subparagraph}[1]{\oldsubparagraph{#1}\mbox{}}
\fi

\usepackage{color}
\usepackage{fancyvrb}

\DefineVerbatimEnvironment{Highlighting}{Verbatim}{commandchars=\\\{\}}
\usepackage{framed}
\definecolor{shadecolor}{RGB}{241,243,245}
\newenvironment{Shaded}{\begin{snugshade}}{\end{snugshade}}

\newcommand{\AttributeTok}[1]{\textcolor[rgb]{0.40,0.45,0.13}{#1}}

\newcommand{\CommentTok}[1]{\textcolor[rgb]{0.37,0.37,0.37}{#1}}

\newcommand{\ConstantTok}[1]{\textcolor[rgb]{0.56,0.35,0.01}{#1}}

\newcommand{\DecValTok}[1]{\textcolor[rgb]{0.68,0.00,0.00}{#1}}

\newcommand{\FloatTok}[1]{\textcolor[rgb]{0.68,0.00,0.00}{#1}}
\newcommand{\FunctionTok}[1]{\textcolor[rgb]{0.28,0.35,0.67}{#1}}

\newcommand{\NormalTok}[1]{\textcolor[rgb]{0.00,0.23,0.31}{#1}}

\newcommand{\OtherTok}[1]{\textcolor[rgb]{0.00,0.23,0.31}{#1}}

\newcommand{\SpecialCharTok}[1]{\textcolor[rgb]{0.37,0.37,0.37}{#1}}

\newcommand{\StringTok}[1]{\textcolor[rgb]{0.13,0.47,0.30}{#1}}

\providecommand{\tightlist}{%
  \setlength{\itemsep}{0pt}\setlength{\parskip}{0pt}}\usepackage{longtable,booktabs,array}
\usepackage{calc} 
\usepackage{etoolbox}
\makeatletter
\patchcmd\longtable{\par}{\if@noskipsec\mbox{}\fi\par}{}{}
\makeatother
\IfFileExists{footnotehyper.sty}{\usepackage{footnotehyper}}{\usepackage{footnote}}
\makesavenoteenv{longtable}
\usepackage{graphicx}
\makeatletter
\def\maxwidth{\ifdim\Gin@nat@width>\linewidth\linewidth\else\Gin@nat@width\fi}
\def\maxheight{\ifdim\Gin@nat@height>\textheight\textheight\else\Gin@nat@height\fi}
\makeatother
\setkeys{Gin}{width=\maxwidth,height=\maxheight,keepaspectratio}
\makeatletter
\def\fps@figure{htbp}
\makeatother
\newlength{\cslhangindent}
\newlength{\csllabelwidth}
\newlength{\cslentryspacingunit} 
\newenvironment{CSLReferences}[2] 
 {
  \setlength{\parindent}{0pt}
  \ifodd #1
  \let\oldpar\par
  \def\par{\hangindent=\cslhangindent\oldpar}
  \fi
  \setlength{\parskip}{#2\cslentryspacingunit}
 }%
 {}
\usepackage{calc}

\newcommand{\CSLLeftMargin}[1]{\parbox[t]{\csllabelwidth}{#1}}
\newcommand{\CSLRightInline}[1]{\parbox[t]{\linewidth - \csllabelwidth}{#1}\break}

\usepackage{booktabs}
\usepackage{caption}
\usepackage{longtable}
\usepackage[noblocks]{authblk}

\usepackage{float}
\makeatletter
\@ifpackageloaded{caption}{}{\usepackage{caption}}
\AtBeginDocument{%
\ifdefined\contentsname
  \renewcommand*\contentsname{Table of contents}
\else
  \newcommand\contentsname{Table of contents}
\fi
\ifdefined\listfigurename
  \renewcommand*\listfigurename{List of Figures}
\else
  \newcommand\listfigurename{List of Figures}
\fi
\ifdefined\listtablename
  \renewcommand*\listtablename{List of Tables}
\else
  \newcommand\listtablename{List of Tables}
\fi
\ifdefined\figurename
  \renewcommand*\figurename{Figure}
\else
  \newcommand\figurename{Figure}
\fi
\ifdefined\tablename
  \renewcommand*\tablename{Table}
\else
  \newcommand\tablename{Table}
\fi
}
\@ifpackageloaded{float}{}{\usepackage{float}}
\floatstyle{ruled}
\@ifundefined{c@chapter}{\newfloat{codelisting}{h}{lop}}{\newfloat{codelisting}{h}{lop}[chapter]}
\floatname{codelisting}{Listing}

\makeatother
\makeatletter
\@ifpackageloaded{caption}{}{\usepackage{caption}}
\@ifpackageloaded{subcaption}{}{\usepackage{subcaption}}
\makeatother
\makeatletter
\@ifpackageloaded{tcolorbox}{}{\usepackage[many]{tcolorbox}}
\makeatother
\makeatletter
\@ifundefined{shadecolor}{\definecolor{shadecolor}{rgb}{.97, .97, .97}}
\makeatother
\ifLuaTeX
  \usepackage{selnolig}  
\fi
\IfFileExists{bookmark.sty}{\usepackage{bookmark}}{\usepackage{hyperref}}
\IfFileExists{xurl.sty}{\usepackage{xurl}}{} 
\hypersetup{
  pdftitle={A modern approach to transition analysis and process mining with Markov models: A tutorial with R},
  pdfauthor={Jouni Helske; Satu Helske; Mohammed Saqr; Sonsoles López-Pernas; Keefe Murphy},
  colorlinks=true,
  linkcolor={blue},
  filecolor={Maroon},
  citecolor={Blue},
  urlcolor={Blue},
  pdfcreator={LaTeX via pandoc}}

\title{A modern approach to transition analysis and process mining with
Markov models: A tutorial with R}

\author[1,*]{Jouni Helske}
\author[2]{Satu Helske}
\author[3]{Mohammed Saqr}
\author[3]{Sonsoles López-Pernas}
\author[4]{Keefe Murphy}

\affil[1]{Department of Mathematics and Statistics, University of
Jyväskylä}
\affil[2]{INVEST Research Flagship Center \& Department of Social
Research, University of Turku}
\affil[3]{School of Computing, University of Eastern Finland}
\affil[4]{Department of Mathematics and Statistics, Hamilton Institute,
Maynooth University}
\affil[*]{Corresponding author: Jouni Helske,
\texttt{jouni.helske@jyu.fi}}

\date{}
\begin{document}
\maketitle
\begin{abstract}
This chapter presents an introduction to Markovian modeling for the
analysis of sequence data. Contrary to the deterministic approach seen
in the previous sequence analysis chapters, Markovian models are
probabilistic models, focusing on the transitions between states instead
of studying sequences as a whole. The chapter provides an introduction
to this method and differentiates between its most common variations:
first-order Markov models, hidden Markov models, mixture Markov models,
and mixture hidden Markov models. In addition to a thorough explanation
and contextualization within the existing literature, the chapter
provides a step-by-step tutorial on how to implement each type of
Markovian model using the R package \texttt{seqHMM}. The chaper also
provides a complete guide to performing stochastic process mining with
Markovian models as well as plotting, comparing and clustering different
process models.
\end{abstract}
\ifdefined\Shaded\renewenvironment{Shaded}{\begin{tcolorbox}[borderline west={3pt}{0pt}{shadecolor}, enhanced, interior hidden, breakable, sharp corners, frame hidden, boxrule=0pt]}{\end{tcolorbox}}\fi

\hypertarget{introduction}{%
\section{Introduction}\label{introduction}}

In the previous two chapters, we have learned about sequence analysis
{[}1, 2{]} and its relevance to educational research. This chapter
delves into a closely-related method: Markovian models. Specifically, we
focus on a particular type of Markovian model, where the data are
assumed to be categorical and observed at discrete time intervals, as
per the previous chapters about sequence analysis, although in general
Markovian models are not restricted to categorical data. One of the main
differences between sequence analysis and Markovian modelling is that
the former relies on deterministic data mining, whereas the latter uses
probabilistic models {[}3{]}. Moreover, sequence analysis takes a more
holistic approach by analysing sequences as a whole, whereas Markovian
modelling focuses on the transitions between states, their probability,
and the reasons (covariates) which explain why these transitions happen.

We provide an introduction and practical guide to the topic of Markovian
models for the analysis of sequence data. While we try to avoid advanced
mathematical notations, to allow the reader to continue to other, more
advanced sources when necessary, we do introduce the basic mathematical
concepts of Markovian models. When doing so, we use the same notation as
in the R package \texttt{seqHMM} {[}4{]}, which we also use in the
examples. In particular, we illustrate first-order Markov models, hidden
Markov models, mixture Markov models, and mixture hidden Markov models
with applications to synthetic data on students' collaboration roles
throughout a complete study program. The Chapter proceeds to describe
the theoretical underpinnings on each method in turn, then showcases
each method with code, before presenting some conclusions and further
readings.

In addition to the aforementioned applications to collaboration roles
and achievement sequences, we also provide a demonstration of the
utility of Markovian models in another context, namely process mining.
In the process mining application, we leverage Markov models and mixture
Markov models to explore learning management system logs. Finally, we
conclude with a brief discussion of Markovian models in general and
provide some recommendations for further reading of advanced topics in
this area as a whole.

\hypertarget{methodological-background}{%
\section{Methodological Background}\label{methodological-background}}

\hypertarget{markov-model}{%
\subsection{Markov model}\label{markov-model}}

\begin{figure}[H]

{\centering \includegraphics[width=1\textwidth,height=\textheight]{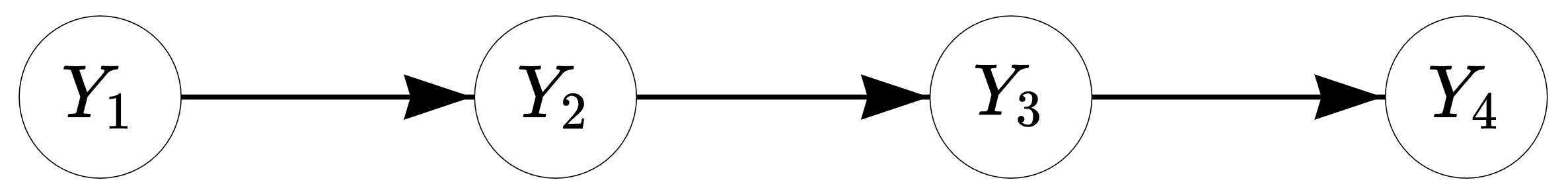}

}

\caption{\label{fig-dag}Illustration of the MM. The quantities \(Y_1\)
to \(Y_4\) refer to states at time points 1 to 4. The arrows indicate
dependencies between states.}

\end{figure}

The simple first-order Markov chain or Markov model (MM) can be used to
model transitions between successive states. In the MM, given the
current observation, the next observation in the sequence is independent
of the past ---this is called the \emph{Markov property}. For example,
when predicting a student's school success in the fourth year we only
need to consider their success in the third year, while their success in
the first and second year give no additional information for the
prediction (see Figure~\ref{fig-dag} for an illustration). As such, the
model is said to be memoryless.

As an example, consider the data described in Table~\ref{tbl-sequences}
which includes four sequences of length ten consisting of an alphabet of
two types of observed state, low achievement success (\(L\)) and high
achievement (\(H\)). Here, the individuals are assumed to be independent
from one another:

\hypertarget{tbl-sequences}{}
\begin{longtable}[]{@{}lllllllllll@{}}
\caption{\label{tbl-sequences}Four example sequences of school
achievement with individuals A--D across the rows and years 1--10 across
the columns.}\tabularnewline
\toprule()
& 1 & 2 & 3 & 4 & 5 & 6 & 7 & 8 & 9 & 10 \\
\midrule()
\endfirsthead
\toprule()
& 1 & 2 & 3 & 4 & 5 & 6 & 7 & 8 & 9 & 10 \\
\midrule()
\endhead
A & L & L & L & H & L & H & L & H & H & H \\
B & L & H & H & L & H & L & H & L & L & H \\
C & H & H & L & H & L & L & H & L & H & H \\
D & H & H & L & L & L & H & L & L & L & H \\
\bottomrule()
\end{longtable}

Say \(t\) describes the position in the sequence, or in this example,
the year (in other words, here \(t\) runs from 1 to 10). If we assume
that the probability of observing \(L\) or \(H\) at any given point
\(t\) depends on the current observation only, we can estimate the
\emph{transition probabilities} \(a_{LL}\) (from state \(L\) to state
\(L\)), \(a_{LH}\) (\(L\) to \(H\)), \(a_{HL}\) (\(H\) to \(L\)), and
\(a_{HH}\) (\(H\) to \(H\)) by calculating the number of observed
transitions from each state to all states and scaling these with the
total number of transitions from that state. Mathematically, we can
write the transition probability \(a_{rs}\) from state \(r\) to state
\(s\) as

\[a_{rs}=P(z_t = s\,|\,z_{t-1} = r), \ s,r \in \{L,H\},\]

which simply states that the observed state \(z_t\) in year \(t\) being
\(L\) or \(H\) depends on which of the two states were observed in the
previous year \(t-1\). For example, to compute
\(a_{LH}=P(z_t=H\,|\,z_{t-1}=L)\), the probability of transitioning from
the origin state \(L\) to the destination state \(H\), we divide the
eight observed transitions to state \(H\) from state \(L\) by 20, which
is the total number of transitions from \(L\) to any state.

The basic MM assumes that the transition probabilities remain constant
in time (this property is called \emph{time-homogeneity}). This means,
for example, that the probabilities of transitioning from the
low-achievement state to the high-achievement state is the same in the
ninth year as it was in the second year. We can collect the transition
probabilities in a transition matrix (which we call \(A\)) which shows
all of the possible transition probabilities between each pair of origin
and destination states, as illustrated in Table~\ref{tbl-transmat}. For
example, when a student has low achievement in year \(t\), they have a
40 percent probability to have low achievement in year \(t+1\) and a
higher 60 percent probability to transition to high achievement instead,
regardless of the year \(t\). Notice that the probabilities in each row
must add up to 1 (or 100\%).

\hypertarget{tbl-transmat}{}
\begin{longtable}[]{@{}lll@{}}
\caption{\label{tbl-transmat}Transition matrix showing the probabilities
of transitioning from one state to another (low or high achievement).
The rows and columns describe the origin state and the destination
state, respectively.}\tabularnewline
\toprule()
& \(\to\) Low & \(\to\) High \\
\midrule()
\endfirsthead
\toprule()
& \(\to\) Low & \(\to\) High \\
\midrule()
\endhead
Low \(\to\) & 8/20 = 0.4 & 12/20 = 0.6 \\
High \(\to\) & 10/16 = 0.625 & 6/16 = 0.375 \\
\bottomrule()
\end{longtable}

Lastly, we need to define probabilities for the starting states of the
sequences, i.e., the \emph{initial probabilities}
\[\pi_s=P(z_1 = s), \ s \in \{L,H\}.\]

In the example, half of the students have low achievement and the other
half have high achievement in the first year, so \(\pi_L=\pi_H = 0.5\).
This basic MM is very simple and is often not realistic in the context
of educational sciences. We can, however, extend the basic MM in several
ways.

First of all, we can include covariates to explain the transition and/or
initial probabilities. For example, if we think that transitioning from
low to high achievement gets harder as the students get older we may add
time as an explanatory variable to the model, allowing the probability
of transitioning from low to high achievement to diminish in time. We
could also increase the order of the Markov chain, accounting for longer
histories. This may be more realistic, but at the same time increasing
the order makes the model considerably more complex, the more so the
longer the history considered.

Secondly, one of the most usable extensions is the inclusion of hidden
(or latent) states that cannot be observed directly but can be estimated
from the sequence of observed states. An MM with time-constant hidden
states is typically called the mixture Markov model (MMM). It can be
used to find latent subpopulations, or in other words, to cluster
sequence data. A model with time-varying hidden states is called the
hidden Markov model (HMM), which allows the individuals to move between
the hidden states. Allowing for both time-constant and time-varying
hidden states leads to a mixture hidden Markov model (MHMM). Unless
otherwise specified, from now on when talking about hidden states we
refer always to time-varying hidden states, while time-constant hidden
states are referred to as clusters.

\hypertarget{mixture-markov-model}{%
\subsection{Mixture Markov model}\label{mixture-markov-model}}

Consider a common case in sequence analysis where individual sequences
are assumed to be clustered to subpopulations such as those with
typically high and low achievement. In the introductory sequence
analysis chapter, the clustering of sequences was performed based on a
matrix of pairwise dissimilarities between sequences. Alternatively, we
can use the MMM to group the sequences based on their initial and
transition probabilities, for example, into those who tend to stay in
and transition to high achievement states and those that tend to stay in
and transition to low achievement states, as illustrated in
Table~\ref{tbl-transmatMMM}.

\begin{table}

\caption{\label{tbl-transmatMMM}Two transition matrices showing the
probabilities of transitioning from one state of achievement to another
in two clusters of Low achievement and High achievement. The rows and
columns describe the origin state and the destination state,
respectively.}\begin{minipage}[t]{0.50\linewidth}
\subcaption{\label{tbl-trans-first}Low achievement}

{\centering 

\begin{tabular}[t]{lll}
\toprule
Cluster: Low achievement & \(\to\) Low & \(\to\) High\\
\midrule
Low \(\to\) & 0.8 & 0.2\\
High \(\to\) & 0.4 & 0.6\\
\bottomrule
\end{tabular}

}

\end{minipage}%
\begin{minipage}[t]{0.50\linewidth}
\subcaption{\label{tbl-trans-second}High achievement}

{\centering 

\begin{tabular}[t]{lll}
\toprule
Cluster: High achievement & \(\to\) Low & \(\to\) High\\
\midrule
Low \(\to\) & 0.6 & 0.4\\
High \(\to\) & 0.1 & 0.9\\
\bottomrule
\end{tabular}

}

\end{minipage}%

\end{table}

In MMMs, we have a separate transition matrix \(A^k\) for each cluster
\(k\) (for \(k=1,\ldots,K\) clusters/subpopulations), and the initial
state distribution defines the probabilities to start (and stay) in the
hidden states corresponding to a particular cluster. This probabilistic
clustering provides group membership probabilities for each sequence;
these define how likely it is that each individual is a member of each
cluster. We can easily add (time-constant) covariates to the model to
explain the probabilities of belonging to each cluster. By incorporating
covariates in this way we could, for example, find that being in a
high-achievement cluster is predicted by gender or family background.
However, we note that this is distinct from the aforementioned potential
inclusion of covariates to explain the transition and/or initial
probabilities.

An advantage of this kind of probabilistic modelling approach is that we
can use traditional model selection methods such as likelihood
information criteria or cross-validation for choosing the best model.
For example, if the number of subpopulations is not known in advance
---as is typically the case--- we can compare models with different
clustering solutions (e.g., those obtained with different numbers of
clusters, different subsets of covariates, or different sets of initial
probabilities, for example) and choose the best-fitting model with, for
example, the Bayesian information criterion (BIC) {[}5{]}.

\hypertarget{hidden-markov-model}{%
\subsection{Hidden Markov model}\label{hidden-markov-model}}

The HMM can be useful in a number of cases when the state of interest
cannot be directly measured or when there is measurement error in the
observations. In HMMs, the Markov chain operates at the level of hidden
states, which subsequently generate or emit observed states with
different probabilities. For example, think about a progression of a
student's ability as a hidden state and school success as the observed
state. We cannot measure true ability directly, but we can estimate the
student's progress by their test scores that are emissions of their
ability. There is, however, some uncertainty in how well the test scores
represent students' true ability. For example, observing low test scores
at some point in time does not necessarily mean the student has low
ability; they might have scored lower than expected in the test due to
other reasons such as being sick at that particular time. Such
uncertainty can be reflected in the emission probabilities; for example,
in the high-ability state students get high test scores eight times out
of ten and low test scores with a 20 percent probability, while in the
low-ability state the students get low test scores nine times out of ten
and high test scores with a 10 percent probability. These probabilities
are collected in an emission matrix as illustrated in
Table~\ref{tbl-emissmatHMM}.

\hypertarget{tbl-emissmatHMM}{}
\begin{longtable}[]{@{}lll@{}}
\caption{\label{tbl-emissmatHMM}Emission matrix showing the
probabilities of each hidden state (low or high ability) emitting each
observed state (low or high test scores).}\tabularnewline
\toprule()
& Low scores & High scores \\
\midrule()
\endfirsthead
\toprule()
& Low scores & High scores \\
\midrule()
\endhead
Low ability & 0.9 & 0.1 \\
High ability & 0.2 & 0.8 \\
\bottomrule()
\end{longtable}

Again, the full HMM is defined by a set of parameters: the initial state
probabilities \(\pi_s\), the hidden state transition probabilities
\(a_{rs}\), and the emission probabilities of observed states
\(b_s(m)\). What is different to the MM is that in the HMM, the initial
state probabilities \(\pi_s\) define the probabilities of starting from
each \emph{hidden} state. Similarly, the transition probabilities
\(a_{rs}\) define the probabilities of transitioning from one
\emph{hidden} state to another hidden state. The emission probabilities
\(b_s(m)\) (collected in an emission matrix \(B\)) define the
probability of observing a particular state \(m\) (e.g., low or high
test scores) given the current hidden state \(s\) (e.g., low or high
ability).

\begin{figure}

{\centering \includegraphics{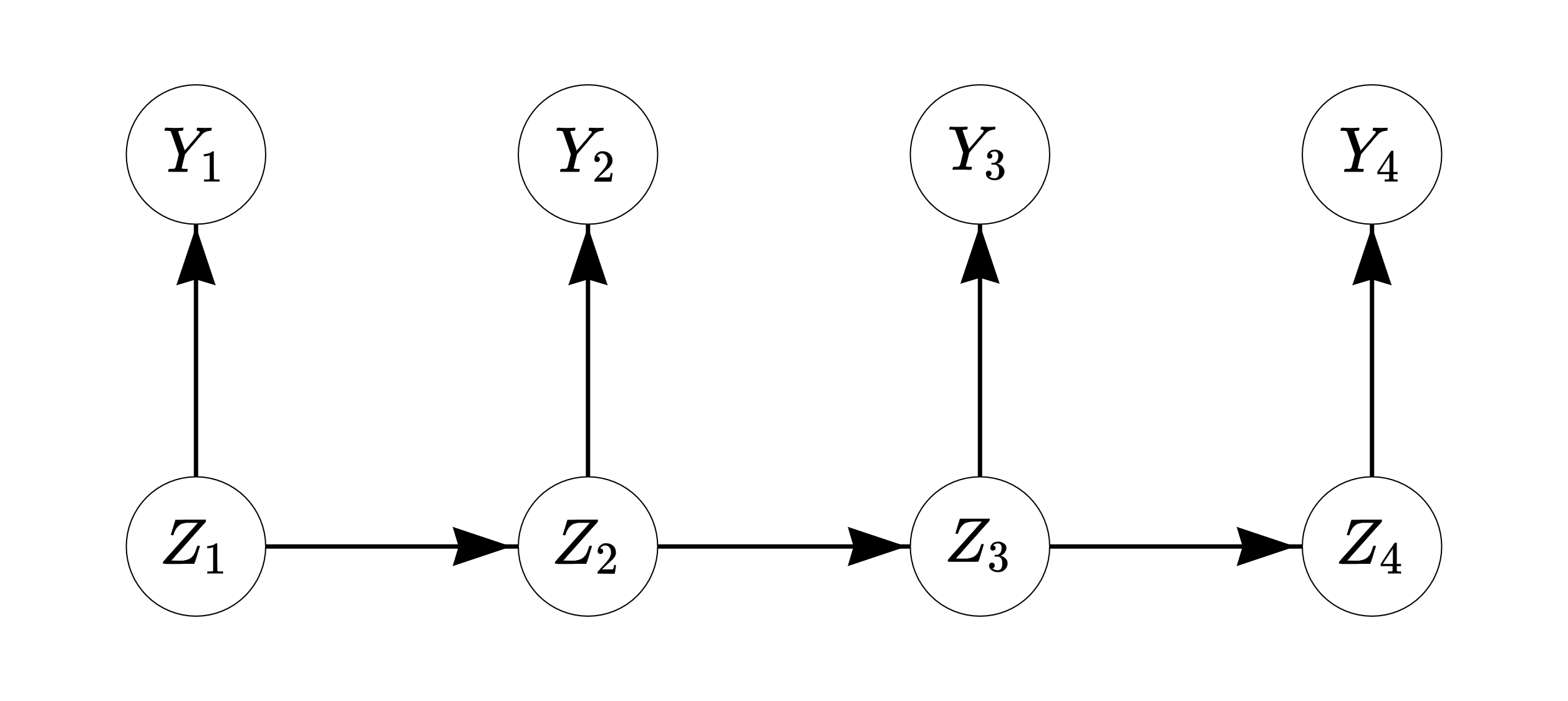}

}

\caption{\label{fig-hmm}Illustration of the HMM. The quantities \(Z_1\)
to \(Z_4\) refer to hidden states at time points 1 to 4, while the
quantities \(Y_1\) to \(Y_4\) refer to observed states. The arrows
indicate dependencies between hidden and/or observed states.}

\end{figure}

When being in a certain hidden state, observed states occur randomly,
following the emission probabilities. Mathematically speaking, instead
of assuming the Markov property directly on our observations, we assume
that the observations are conditionally independent given the underlying
hidden state. We can visualise the HMM as a directed acyclic graph (DAG)
illustrated in Figure~\ref{fig-hmm}. Here \(Z\) are the unobserved
states (such as ability) which affect the distribution of the observed
states \(Y\) (test scores). At each time point \(t\), the state \(z_t\)
can obtain one of \(S\) possible values (there are two hidden states in
the example of low and high ability, so \(S=2\)), which in turn defines
how \(Y_t\) is distributed.

\hypertarget{mixture-hidden-markov-models}{%
\subsection{Mixture hidden Markov
models}\label{mixture-hidden-markov-models}}

Combining the ideas of both time-constant clusters and time-varying
hidden states leads to the concept of mixture hidden Markov model
(MHMM). Here we assume that the population of interest consists of
subpopulations, each with their own HMM with varying transition and
emission probabilities. For example, we could expect to find underlying
groups which behave differently when estimating the progression of
ability through the sequence of test scores, such as those that
consistently stay on a low-ability or high-ability track (stayers) and
those that move between low and high ability (movers). In this case, we
need two transition matrices: the stayers' transition matrix allows for
no transitions while the movers' transition matrix allows for
transitioning between low and high ability, as illustrated in
Table~\ref{tbl-transmatMHMM}.

\begin{table}

\caption{\label{tbl-transmatMHMM}Two transition matrices showing the
probabilities of transitioning from one state of ability to another in
two clusters, the Stayers and the Movers. The rows and columns describe
the origin state and the destination state,
respectively.}\begin{minipage}[t]{0.50\linewidth}
\subcaption{\label{tbl-transmatMHMM-stayers}Stayers}

{\centering 

\begin{tabular}[t]{lll}
\toprule
Cluster: Stayers & \(\to\) Low & \(\to\) High\\
\midrule
Low \(\to\) & 1 & 0\\
High \(\to\) & 0 & 1\\
\bottomrule
\end{tabular}

}

\end{minipage}%
\begin{minipage}[t]{0.50\linewidth}
\subcaption{\label{tbl-transmatMHMM-movers}Movers}

{\centering 

\begin{tabular}[t]{lll}
\toprule
Cluster: Movers & \(\to\) Low & \(\to\) High\\
\midrule
Low \(\to\) & 0.6 & 0.4\\
High \(\to\) & 0.3 & 0.7\\
\bottomrule
\end{tabular}

}

\end{minipage}%

\end{table}

Similarly, we need two emission matrices that describe how the observed
states are related to hidden states, as illustrated in
Table~\ref{tbl-emissmatMHMM}. In this example, there is a closer match
between low/high ability and low/high test scores in the Stayers cluster
in comparison to the Movers cluster.

\begin{table}

\caption{\label{tbl-emissmatMHMM}Two emission matrices showing the
probabilities of each hidden state (low or high ability) emitting each
observed state (low or high test
scores).}\begin{minipage}[t]{0.50\linewidth}
\subcaption{\label{tbl-emissmatMHMM-stayers}Stayers}

{\centering 

\begin{tabular}[t]{lll}
\toprule
Cluster: Stayers & Low scores & High scores\\
\midrule
Low ability & 0.9 & 0.1\\
High ability & 0.1 & 0.9\\
\bottomrule
\end{tabular}

}

\end{minipage}%
\begin{minipage}[t]{0.50\linewidth}
\subcaption{\label{tbl-emissmatMHMM-movers}Movers}

{\centering 

\begin{tabular}[t]{lll}
\toprule
Cluster: Movers & Low scores & High scores\\
\midrule
Low ability & 0.7 & 0.3\\
High ability & 0.2 & 0.8\\
\bottomrule
\end{tabular}

}

\end{minipage}%

\end{table}

Mathematically, when estimating a MHMM we first fix the number of
clusters \(K\), and create a joint HMM consisting of \(K\) submodels
(HMMs). The number of hidden states does not have to be fixed but can
vary by submodel, so that the HMMs have more hidden states for some
clusters and fewer for others (in our example, because the transition
matrix is of the Stayers cluster is diagonal, we could also split the
cluster into two single state clusters, one corresponding to low and
other to high ability). This can increase the burden of model selection,
so often a common number of hidden states is assumed for each cluster
for simplicity. In any case, the initial state probabilities of this
joint model define how sequences are assigned to different clusters. We
estimate this joint model using the whole data and calculate cluster
membership probabilities for each individual. The idea of using mixtures
of HMMs has appeared in literature under various names with slight
variations, e.g., {[}6{]}, {[}7{]}, and {[}4{]}. Notably, MHMMs inherit
from MMMs the ability to incoporate covariates to predict cluster
memberships.

\hypertarget{multi-channel-sequences}{%
\subsection{Multi-channel sequences}\label{multi-channel-sequences}}

There are two options to analyse multi-channel (or multi-domain or
multi-dimensional) sequence data with Markovian models. The first option
is to combine observed states in different channels into one set of
single-channel sequences with an expanded alphabet. This option is
simple, and works for MMs, HMMs, MMMs, and MHMMs, but can easily lead to
complex models as the number of states and channels increases
considerably. The second option, which can only be used when working
with HMMs and MHMMs, is to treat the observed states in each channel
independently given the current hidden state. This can be easily
performed by defining multiple emission probability matrices, one for
each channel. The assumption of conditional independence simplifies the
model, but is sometimes unrealistic, in which case it is better to
resort to the first option and convert the data into single-channel
sequences. Both options are discussed further in Chapter 13 {[}8{]}, a
dedicated chapter on multi-channel sequences, where applications of
distance-based and Markovian clustering approaches are presented. In
this chapter, we henceforth focus on single-channel sequences.

\hypertarget{estimating-model-parameters}{%
\subsection{Estimating model
parameters}\label{estimating-model-parameters}}

The model parameters, i.e.~the elements of the initial probability
vectors \(\pi\), transition probability matrices \(A\), and emission
probability matrices \(B\), can be estimated from data using various
methods. Typical choices are the Baum-Welch algorithm (an instance of
the expectation-maximisation, i.e., the EM algorithm) and direct
(numerical) maximum likelihood estimation. It is possible to restrict
models, for example, by setting some parameters to fixed values
(typically zeros), for example, to make certain starting states,
transitions, or emissions impossible.

After the parameter estimation, in addition to studying the estimated
model parameters upon convergence, we can, for example, compute
cluster-membership probabilities for each individual and find the most
probable paths of hidden state sequences using the Viterbi algorithm
({[}9{]}). These can be further analysed and visualised for
interpretation.

\hypertarget{review-of-the-literature}{%
\section{Review of the literature}\label{review-of-the-literature}}

Markovian methods have been used across several domains in education and
have gained renewed interest with the surge in learning analytics and
educational data mining. Furthermore, the introduction of specialised R
packages (e.g., \texttt{seqHMM} {[}10{]}) and software applications
(e.g., Mplus {[}11, 12{]}) have made it easier to implement Markov
models. One of the most common applications of Markovian methods is the
clustering of sequence data {[}13--15{]}. Markov models offer a credible
alternative to existing distance-based methods (e.g.~optimal matching)
and can be used with different sequence types (e.g.~multi-channel
sequences). Furthermore, Markovian methods offer some advantages in
clustering sequential data such as the inclusion of covariates that can
explain why a sequence emerged (e.g., {[}16{]}). More importantly,
Markovian models are relatively scalable and can be used to cluster
large sequences {[}17{]}. As Saqr et al. {[}17{]} noted, large sequences
are hard to cluster using standard methods such as hierarchical
clustering, which is memory inefficient, and hard to parallelise or
scale {[}18, 19{]}. Furthermore, distance-based clustering methods are
limited by the theoretical maximum dimension of a matrix in R which is
2,147,483,647 (i.e., a maximum of 46,430 sequences). In such a case,
Markovian methods may be the solution.

Examples of Markovian methods in clustering sequences are plentiful. For
example, HMMs have been used to cluster students' sequences of learning
management system (LMS) trace data to detect their patterns of
activities or what the authors referred to as learning tactics and
strategies {[}15{]}. Another close example was that of López-Pernas and
Saqr {[}20{]}, who used HMMs to cluster multi-channel data of students'
learning strategies of two different tools (an LMS and an automated
assessment tool). Other examples include using HMM in clustering
sequences of students' engagement states {[}21{]}, sequences of
students' collaborative roles {[}16{]}, or sequences of self-regulation
{[}13, 14{]}.

Markovian methods are also popular in studying transitions and have
therefore been used across several applications and with different types
of data. One of the most common usages is what is known as stochastic
processes mining which typically uses first-order Markov models to map
students' transitions between learning activities. For example, Matcha
et al. {[}22{]} used first-order Markov models to study students'
processes of transitions between different learning tactics. Other uses
include studying the transitions between tactics of academic writing
{[}23{]}, between self-regulated learning events {[}24{]}, or within
collaborative learning settings {[}25{]}. Yet, most of such work has
been performed by the \texttt{pMiner} R package {[}26{]}, which was
recently removed from The Comprehensive R Archive Network (CRAN) due to
slow updates and incompatibility with existing guidelines. This chapter
offers a modern alternative that uses modern and flexible methods for
fitting, plotting, and clustering stochastic process mining models as
well as the possibility to add covariates to understand ``why''
different transitions pattern emerged.

Indeed, transition analysis in general has been a popular usage for
Markovian models and has been used across several studies. For instance,
for the analysis of temporal patterns of students' activities in online
learning (e.g., {[}27{]}), or transitions between latent states
{[}28{]}, or transitions between assignment submission patterns
{[}29{]}.

\hypertarget{examples}{%
\section{Examples}\label{examples}}

As a first step, we will import all the packages required for our
analyses. We have used most of them throughout the book. Below is a
brief summary:

\begin{itemize}
\tightlist
\item
  \texttt{qgraph}: A package for visualising networks, which can be used
  to plot transition probabilities {[}30{]}. This is used only for the
  process mining application in Section \ref{process}.
\item
  \texttt{rio}: A package for reading and saving data files with
  different extensions {[}31{]}.
\item
  \texttt{seqHMM}: A package designed for fitting hidden (latent) Markov
  models and mixture hidden Markov models for social sequence data and
  other categorical time series {[}32{]}.
\item
  \texttt{tidyverse}: A package that encompasses several basic packages
  for data manipulation and wrangling {[}33{]}.
\item
  \texttt{TraMineR}: As seen in the introductory sequence analysis
  chapter, this package helps us construct, analyze, and visualise
  sequences from time-ordered states or events {[}34{]}.
\end{itemize}

\begin{Shaded}
\begin{Highlighting}[]
\FunctionTok{library}\NormalTok{(qgraph)}
\FunctionTok{library}\NormalTok{(rio)}
\FunctionTok{library}\NormalTok{(seqHMM)}
\FunctionTok{library}\NormalTok{(tidyverse)}
\FunctionTok{library}\NormalTok{(TraMineR)}
\end{Highlighting}
\end{Shaded}

Henceforth, we divide our examples into two parts: the first largely
focuses on traditional uses of the \texttt{seqHMM} package to fit
Markovian models of varying complexity to sequence data; the latter
presents a demonstration of Markovian models from the perspective of
process mining. We outline the steps involved in using \texttt{seqHMM}
in general in Section \ref{steps}, demonstrate the application of MMs,
HMMs, MMMs, and MHMMs in Section \ref{markov}, and explore process
mining using Markovian models in Section \ref{process}, leveraging much
of the steps and code from the previous two sections. We note that
different datasets are used in Section \ref{markov} and Section
\ref{process}; we begin by importing the data required for Section
\ref{markov} and defer the importing of the data used in the process
mining application to the later section.

With this in mind, we start by using the \texttt{ìmport()} function from
the \texttt{rio} package to import our sequence data. Based on the
description of the MHMM in {[}35{]}, we used the \texttt{seqHMM} package
to simulate a synthetic dataset (\texttt{simulated\_data}) consisting of
students' collaboration roles (obtained from {[}36{]}) on different
courses across a whole program. As the original data, the simulation was
based on the two-channel model (collaboration and achievement), but we
only use the collaboration sequences in the following examples, and
leave the multi-channel sequence analysis to Chapter 13 {[}8{]}. While
not available in the original study, we also simulated students' high
school grade point average (\texttt{GPA}, for simplicity categorised
into three levels) for each student, which will be used to predict
cluster memberships. Using this data, we show how the \texttt{seqHMM}
package can be used to analyse such sequences. We start with the simple
MM, and then transition to HMMs and their mixtures. To be able to use
the \texttt{seqHMM} functions we need to convert the imported data to a
sequence using the function \texttt{seqdef} from the \texttt{TraMineR}
package (see Chapter 10 {[}1{]} for more information about creating
\texttt{stslist} objects). We can also extract the covariate information
separately (\texttt{cov\_data}).

\begin{Shaded}
\begin{Highlighting}[]
\NormalTok{URL }\OtherTok{\textless{}{-}} \StringTok{"https://github.com/sonsoleslp/labook{-}data/raw/main/"}
\NormalTok{simulated\_data }\OtherTok{\textless{}{-}} \FunctionTok{import}\NormalTok{(}\FunctionTok{paste0}\NormalTok{(URL, }\StringTok{"12\_longitudinalRoles/simulated\_roles.csv"}\NormalTok{))}

\NormalTok{roles\_seq }\OtherTok{\textless{}{-}} \FunctionTok{seqdef}\NormalTok{(simulated\_data, }\AttributeTok{var =} \DecValTok{3}\SpecialCharTok{:}\DecValTok{22}\NormalTok{, }\AttributeTok{alphabet =} \FunctionTok{c}\NormalTok{(}\StringTok{"Isolate"}\NormalTok{, }\StringTok{"Mediator"}\NormalTok{, }\StringTok{"Leader"}\NormalTok{),}
          \AttributeTok{cnames =} \DecValTok{1}\SpecialCharTok{:}\DecValTok{20}\NormalTok{)}
\end{Highlighting}
\end{Shaded}

\begin{verbatim}
 [>] 3 distinct states appear in the data: 
\end{verbatim}

\begin{verbatim}
     1 = Isolate
\end{verbatim}

\begin{verbatim}
     2 = Leader
\end{verbatim}

\begin{verbatim}
     3 = Mediator
\end{verbatim}

\begin{verbatim}
 [>] state coding:
\end{verbatim}

\begin{verbatim}
       [alphabet]  [label]  [long label] 
\end{verbatim}

\begin{verbatim}
     1  Isolate     Isolate  Isolate
\end{verbatim}

\begin{verbatim}
     2  Mediator    Mediator Mediator
\end{verbatim}

\begin{verbatim}
     3  Leader      Leader   Leader
\end{verbatim}

\begin{verbatim}
 [>] 200 sequences in the data set
\end{verbatim}

\begin{verbatim}
 [>] min/max sequence length: 20/20
\end{verbatim}

\begin{Shaded}
\begin{Highlighting}[]
\FunctionTok{cpal}\NormalTok{(roles\_seq) }\OtherTok{\textless{}{-}} \FunctionTok{c}\NormalTok{(}\StringTok{"\#FBCE4B"}\NormalTok{, }\StringTok{"\#F67067"}\NormalTok{, }\StringTok{"\#5C2262"}\NormalTok{)}

\NormalTok{cov\_data }\OtherTok{\textless{}{-}}\NormalTok{ simulated\_data }\SpecialCharTok{\%\textgreater{}\%} 
  \FunctionTok{select}\NormalTok{(ID, GPA) }\SpecialCharTok{\%\textgreater{}\%} 
  \FunctionTok{mutate}\NormalTok{(}\AttributeTok{GPA =} \FunctionTok{factor}\NormalTok{(GPA, }\AttributeTok{levels =} \FunctionTok{c}\NormalTok{(}\StringTok{"Low"}\NormalTok{, }\StringTok{"Middle"}\NormalTok{, }\StringTok{"High"}\NormalTok{)))}
\end{Highlighting}
\end{Shaded}

\hypertarget{steps}{%
\subsection{Steps of estimation}\label{steps}}

We will first briefly introduce the steps of the analysis with the
\texttt{seqHMM} package and then show examples of estimating MMs, HMMs,
MMMs, and MHMMs.

\hypertarget{defining-the-model-structure}{%
\subsubsection{Defining the model
structure}\label{defining-the-model-structure}}

First, we need to create the model object which defines the structure of
the model. This can be done by using one of the model building functions
of \texttt{seqHMM}. The build functions include \texttt{build\_mm()} for
constructing the simple MM, \texttt{build\_hmm()} for the HMM,
\texttt{build\_mmm()} for the MMM, and \texttt{build\_mhmm()} for the
MHMM. The user needs to give the build function the sequence data and
the number of hidden states and/or clusters (when relevant). The user
can also set restrictions on the models, for example, to forbid some
transitions by setting the corresponding transition probabilities to
zero. To facilitate the estimation of the parameters of more complex
models, the user may also set informative starting values for model
parameters.

\hypertarget{estimating-the-model-parameters}{%
\subsubsection{Estimating the model
parameters}\label{estimating-the-model-parameters}}

After defining the model structure, model parameters need to be
estimated. The \texttt{fit\_model()} function estimates model parameters
using maximum likelihood estimation. The function has several arguments
for configuring the estimation algorithms. For simple models the default
arguments tend to work well enough, but for more complex models the user
should adjust the algorithms. This is because the more parameters the
algorithm needs to estimate, the higher the risk of not finding the
model with the optimal parameter values (the one which maximises the
likelihood).

In order to reduce the risk of being trapped in a local optimum of the
likelihood surface (instead of a global optimum), we advise to estimate
the model numerous times using different starting values for the
parameters. The \texttt{seqHMM} package strives to automate this. One
option is to run the EM algorithm multiple times with random starting
values for any or all of initial, transition, and emission
probabilities. These are specified in the \texttt{control\_em} argument.
Although not done by default, this method seems to perform very well as
the EM algorithm is relatively fast. Another option is to use a global
direct numerical estimation method such as the multilevel single-linkage
method. See {[}4{]} for more detailed information on model estimation.

\hypertarget{examining-the-results}{%
\subsubsection{Examining the results}\label{examining-the-results}}

The output of the \texttt{fit\_model} contains the estimated model
(stored in \texttt{fit\_hmm\$model}) as well as information about the
estimation of the model, such as the log-likelihood of the final model
(\texttt{fit\_hmm\$logLik}). The \texttt{print} method provides
information about the estimated model in a written format, while the
\texttt{plot()} function visualises the model parameters as a graph. For
HMMs and MHMMs, we can calculate the most probable sequence of hidden
states for each individual with the \texttt{hidden\_paths()} function.
Sequences of observed and hidden state sequences can be plotted with the
\texttt{ssplot()} function for MMs and HMMs and with the
\texttt{mssplot()} function for the MMMs and the MHMMs. For MMMs and
MHMMs, the \texttt{summary()} method automatically computes some
features of the models such as standard errors for covariates and prior
and posterior cluster membership probabilities for the subjects.

\hypertarget{markov}{%
\subsection{Markov models}\label{markov}}

We now follow the steps outlined above for each model in turn, starting
from the most basic Markov model, proceeding through a hidden Markov
model and a mixture Markov model, and finally concluding with a mixture
hidden Markov model.

\hypertarget{markov-model-1}{%
\subsubsection{Markov model}\label{markov-model-1}}

We focus on the sequences of collaboration roles, collected in the
\texttt{roles\_seq} object. The \texttt{build\_mm()} function only takes
one argument, \texttt{observations}, which should be an \texttt{stslist}
object created with the \texttt{seqdef()} function from the
\texttt{TraMineR} package as mentioned before. We can build a MM as
follows:

\begin{Shaded}
\begin{Highlighting}[]
\NormalTok{markov\_model }\OtherTok{\textless{}{-}} \FunctionTok{build\_mm}\NormalTok{(roles\_seq)}
\end{Highlighting}
\end{Shaded}

For the MM, the \texttt{build\_mm()} function estimates the initial
probabilities and the transition matrix. Note that the
\texttt{build\_mm()} function is the only build function that
automatically estimates the parameters of the model. This is possible
because for the MM the estimation is a simple calculation while for the
other types of models the estimation process is more complex. The user
can access the estimated parameters by calling
\texttt{markov\_model\$initial\_probs} and
\texttt{markov\_model\$transition\_probs} or view them by using the
print method of the model:

\begin{Shaded}
\begin{Highlighting}[]
\FunctionTok{print}\NormalTok{(markov\_model)}
\end{Highlighting}
\end{Shaded}

\begin{verbatim}
Initial probabilities :
 Isolate Mediator   Leader 
   0.375    0.355    0.270 

Transition probabilities :
          to
from       Isolate Mediator Leader
  Isolate   0.4231    0.478 0.0987
  Mediator  0.1900    0.563 0.2467
  Leader    0.0469    0.428 0.5252
\end{verbatim}

We can see that the initial state probabilities are relatively uniform,
with a slightly lower probability for starting in the Leader state. In
terms of the transition probabilities, the most distinct feature is that
that it is rare to transition directly from the Leader state to Isolate
and vice versa (estimated probabilities are about 5\% and 10\%,
respectively). It is also more common to drop from Leader to Mediator
(43\%) than to increase collaboration from Mediator to Leader (25\%).
Similarly, the probability of moving from Mediator to Isolate is only 19
percent, but there is a 48 percent chance of transitioning from Isolate
to Mediator.

\begin{figure}[H]

{\centering \includegraphics{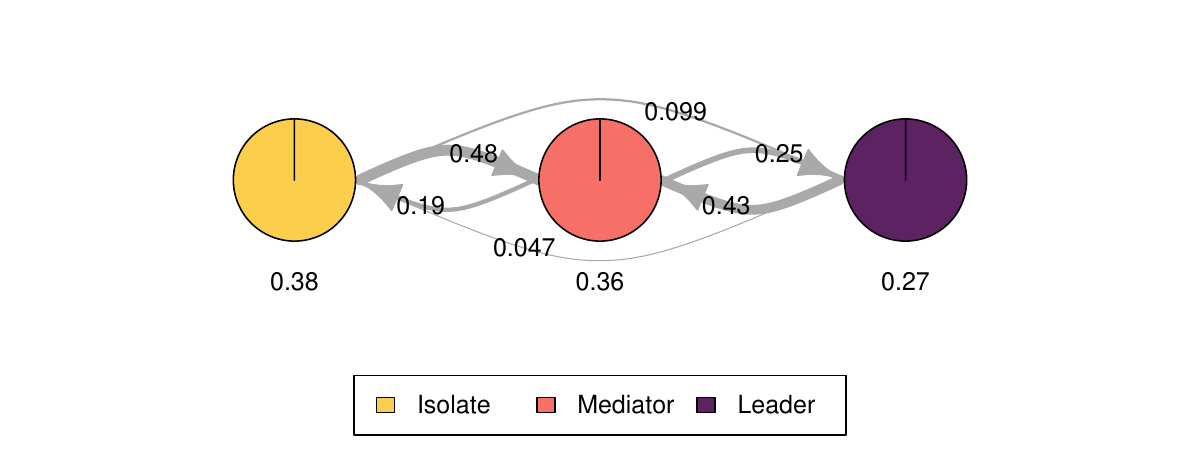}

}

\caption{\label{fig-mm-pie}MM estimated model pie chart.}

\end{figure}

We can also draw a graph of the estimated model using the \texttt{plot}
method which by default shows the states as pie graphs (for the MM, the
pie graphs only consist of one state), transition probabilities as
directed arrows, and initial probabilities below each state (see
Figure~\ref{fig-mm-pie}).

\begin{Shaded}
\begin{Highlighting}[]
\FunctionTok{plot}\NormalTok{(markov\_model, }\AttributeTok{legend.prop =} \FloatTok{0.2}\NormalTok{, }\AttributeTok{ncol.legend =} \DecValTok{3}\NormalTok{, }
     \AttributeTok{edge.label.color =} \StringTok{"black"}\NormalTok{, }\AttributeTok{vertex.label.color =} \StringTok{"black"}\NormalTok{)}
\end{Highlighting}
\end{Shaded}

\hypertarget{hidden-markov-models}{%
\subsubsection{Hidden Markov models}\label{hidden-markov-models}}

The structure of an HMM is set with the \texttt{build\_hmm()} function.
In contrast to \texttt{build\_mm()}, other \texttt{build\_*()} functions
such as \texttt{build\_hmm()} do not directly estimate the model
parameters. For \texttt{build\_hmm()}, in addition to observations (an
\texttt{stslist}), we need to provide the \texttt{n\_states} argument
which tells the model how many hidden states to construct. Using again
the collaboration roles sequences, if we want to estimate an HMM with
two hidden states, we can write:

\begin{Shaded}
\begin{Highlighting}[]
\FunctionTok{set.seed}\NormalTok{(}\DecValTok{1}\NormalTok{)}
\NormalTok{hidden\_markov\_model }\OtherTok{\textless{}{-}} \FunctionTok{build\_hmm}\NormalTok{(}\AttributeTok{observations =}\NormalTok{ roles\_seq, }\AttributeTok{n\_states =} \DecValTok{2}\NormalTok{)}
\end{Highlighting}
\end{Shaded}

The \texttt{set.seed} call ensures that we will always end up with the
same exact initial model with hidden states in the same exact order even
though we use random values for the initial parameters of the model
(which is practical for reproducibility). We are now ready to estimate
the model with the \texttt{fit\_model()} function. The HMM we want to
estimate is simple, so we rely on the default values and again use the
print method to provide information about the estimated model:

\begin{Shaded}
\begin{Highlighting}[]
\NormalTok{fit\_hmm }\OtherTok{\textless{}{-}} \FunctionTok{fit\_model}\NormalTok{(hidden\_markov\_model)}
\NormalTok{fit\_hmm}\SpecialCharTok{$}\NormalTok{model}
\end{Highlighting}
\end{Shaded}

\begin{verbatim}
Initial probabilities :
State 1 State 2 
  0.657   0.343 

Transition probabilities :
         to
from      State 1 State 2
  State 1  0.9089  0.0911
  State 2  0.0391  0.9609

Emission probabilities :
           symbol_names
state_names Isolate Mediator Leader
    State 1  0.4418    0.525 0.0336
    State 2  0.0242    0.478 0.4980
\end{verbatim}

\begin{figure}[H]

{\centering \includegraphics{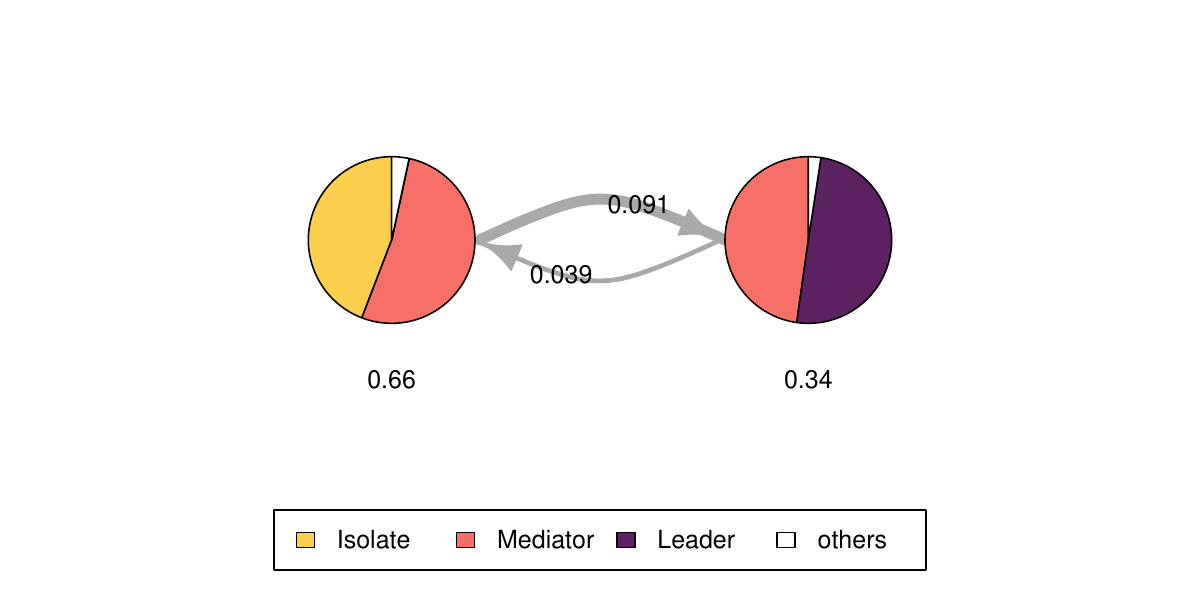}

}

\caption{\label{fig-hmm-pie}HMM with two hidden states.}

\end{figure}

The estimated initial state probabilities show that it is more probable
to start from hidden state 1 than from hidden state 2 (66\% vs.~34\%).
The high transition probabilities on the diagonal of the transition
matrix indicate that the students typically tend to stay in the hidden
state they currently are in. Transition probabilities between the hidden
states are relatively low and also asymmetric: it is more likely that
students move from state 1 to state 2 than from state 2 to state 1.
Looking at the emission matrices, we see that the role of the students
in state 2 is mostly Leader or Mediator (emission probabilities are 50\%
and 48\%). On the other hand, state 1 captures more of those occasions
where students are isolated or exhibit at most a moderate level of
participation (mediators). We can also visualise this with the
\texttt{plot()} method of \texttt{seqHMM} (see
Figure~\ref{fig-hmm-pie}):

\begin{Shaded}
\begin{Highlighting}[]
\FunctionTok{plot}\NormalTok{(fit\_hmm}\SpecialCharTok{$}\NormalTok{model, }\AttributeTok{ncol.legend =} \DecValTok{4}\NormalTok{, }\AttributeTok{legend.prop =} \FloatTok{0.2}\NormalTok{, }
     \AttributeTok{edge.label.color =} \StringTok{"black"}\NormalTok{, }\AttributeTok{vertex.label.color =} \StringTok{"black"}\NormalTok{)}
\end{Highlighting}
\end{Shaded}

The plot values mainly shows the same information. By default, to
simplify the graph, the plotting method combines all states with less
than 5\% emission probabilities into one category. This threshold can be
changed with the \texttt{combine.slices} argument (setting
\texttt{combine.slices\ =\ 0} plots all states).

For simple models, using \texttt{n\_states} is sufficient. It
automatically draws random starting values that are then used for the
estimation of model parameters. However, as parameter estimation of HMMs
and mixture models can be sensitive to starting values of parameters, it
may be beneficial to provide starting values manually using the
\texttt{initial\_probs}, \texttt{transition\_probs}, and
\texttt{emission\_probs} arguments. This is also necessary in case we
want to define structural zeros for some of these components, e.g., if
we want to restrict the initial probabilities so that each sequence
starts from the same hidden state, or if we want to set an upper
diagonal transition matrix, which means that the model does not allow
transitioning back to previous states (this is called a left-to-right
model) {[}4{]}. It is also possible to mix random and user-defined
starting values by using \texttt{simulate\_*()} functions
(e.g.~\texttt{simulate\_transition\_probs()}) for some of the model
components and user-defined values for others.

In the following example we demonstrate estimating a three-state HMM
with user-defined starting values for the initial state probabilities
and the transition matrix but simulate starting values for the emission
matrices. For simulating starting values with
\texttt{simulate\_emission\_probs}, we need to define the number of
hidden states, and the number of observed symbols, i.e., the length of
the alphabet of the sequence data.

\begin{Shaded}
\begin{Highlighting}[]
\CommentTok{\# Set seed for randomisation}
\FunctionTok{set.seed}\NormalTok{(}\DecValTok{1}\NormalTok{)}

\CommentTok{\# Initial state probability vector, must sum to one}
\NormalTok{init\_probs }\OtherTok{\textless{}{-}} \FunctionTok{c}\NormalTok{(}\FloatTok{0.3}\NormalTok{, }\FloatTok{0.4}\NormalTok{, }\FloatTok{0.3}\NormalTok{)}

\CommentTok{\# a 3x3 transition matrix, each row should sum to one}
\NormalTok{trans\_probs }\OtherTok{\textless{}{-}} \FunctionTok{rbind}\NormalTok{(}\FunctionTok{c}\NormalTok{(}\FloatTok{0.8}\NormalTok{, }\FloatTok{0.15}\NormalTok{, }\FloatTok{0.05}\NormalTok{), }\FunctionTok{c}\NormalTok{(}\FloatTok{0.2}\NormalTok{, }\FloatTok{0.6}\NormalTok{, }\FloatTok{0.2}\NormalTok{), }\FunctionTok{c}\NormalTok{(}\FloatTok{0.05}\NormalTok{, }\FloatTok{0.15}\NormalTok{, }\FloatTok{0.8}\NormalTok{))}

\CommentTok{\# Simulate emission probabilities}
\NormalTok{emission\_probs }\OtherTok{\textless{}{-}} \FunctionTok{simulate\_emission\_probs}\NormalTok{(}\AttributeTok{n\_states =} \DecValTok{3}\NormalTok{, }
                                          \AttributeTok{n\_symbols =} \FunctionTok{length}\NormalTok{(}\FunctionTok{alphabet}\NormalTok{(roles\_seq)))}

\CommentTok{\# Build the HMM}
\NormalTok{hidden\_markov\_model\_2 }\OtherTok{\textless{}{-}} \FunctionTok{build\_hmm}\NormalTok{(roles\_seq,}
                                   \AttributeTok{initial\_probs =}\NormalTok{ init\_probs, }
                                   \AttributeTok{transition\_probs =}\NormalTok{ trans\_probs,}
                                   \AttributeTok{emission\_probs =}\NormalTok{ emission\_probs)}
\end{Highlighting}
\end{Shaded}

Our initial probabilities suggest that it is slightly more likely to
start from the second hidden state than the first and the third.
Furthermore, the starting values for the transition matrices suggest
that staying in hidden states 1 and 3 is more likely than staying in
hidden state 2. All non-zero probabilities are, however, mere
suggestions and will be estimated with the \texttt{fit\_model()}
function. We now estimate this model 50 times with the EM algorithm
using randomised starting values:

\begin{Shaded}
\begin{Highlighting}[]
\FunctionTok{set.seed}\NormalTok{(}\DecValTok{1}\NormalTok{)}
\NormalTok{fit\_hmm\_2 }\OtherTok{\textless{}{-}} \FunctionTok{fit\_model}\NormalTok{(hidden\_markov\_model\_2, }
                       \AttributeTok{control\_em =} \FunctionTok{list}\NormalTok{(}\AttributeTok{restart =} \FunctionTok{list}\NormalTok{(}\AttributeTok{times =} \DecValTok{50}\NormalTok{)))}
\end{Highlighting}
\end{Shaded}

We can get the information on the EM estimation as follows:

\begin{Shaded}
\begin{Highlighting}[]
\NormalTok{fit\_hmm\_2}\SpecialCharTok{$}\NormalTok{em\_results}
\end{Highlighting}
\end{Shaded}

\begin{verbatim}
$logLik
[1] -3546.155

$iterations
[1] 488

$change
[1] 9.947132e-11

$best_opt_restart
 [1] -3546.155 -3546.155 -3546.155 -3546.155 -3546.155 -3546.155 -3546.155
 [8] -3546.155 -3546.155 -3546.155 -3546.155 -3546.155 -3546.155 -3546.155
[15] -3546.155 -3546.155 -3546.155 -3546.155 -3546.155 -3546.155 -3546.155
[22] -3546.155 -3546.155 -3546.155 -3546.155
\end{verbatim}

The \texttt{loglik} element gives the log-likelihood of the final model.
This value has no meaning on its own, but it can be used to compare HMMs
with the same data and model structure (e.g., when estimating the same
model from different starting values). The \texttt{iterations} and
\texttt{change} arguments give information on the last EM estimation
round: how many iterations were used until the (local) optimum was found
and what was the change in the log-likelihood at the final step.

The most interesting element is the last one:
\texttt{best\_opt\_restart} shows the likelihood for 25 (by default) of
the best estimation rounds. We advise to always check these to make sure
that the best model was found several times from different starting
values: this way we can be fairly certain that we have found the actual
maximum likelihood estimates of the model parameters (global optimum).
In this case all of the 25 log-likelihood values are identical, meaning
that it is likely that we have found the best possible model among all
HMMs with three hidden states.

\begin{Shaded}
\begin{Highlighting}[]
\FunctionTok{plot}\NormalTok{(fit\_hmm\_2}\SpecialCharTok{$}\NormalTok{model, }\AttributeTok{legend.prop =} \FloatTok{0.15}\NormalTok{, }\AttributeTok{ncol.legend =} \DecValTok{3}\NormalTok{,}
     \AttributeTok{edge.label.color =} \StringTok{"black"}\NormalTok{, }\AttributeTok{vertex.label.color =} \StringTok{"black"}\NormalTok{,}
     \AttributeTok{combine.slices =} \DecValTok{0}\NormalTok{, }\AttributeTok{trim =} \FloatTok{0.0001}\NormalTok{)}
\end{Highlighting}
\end{Shaded}

\begin{figure}[H]

{\centering \includegraphics{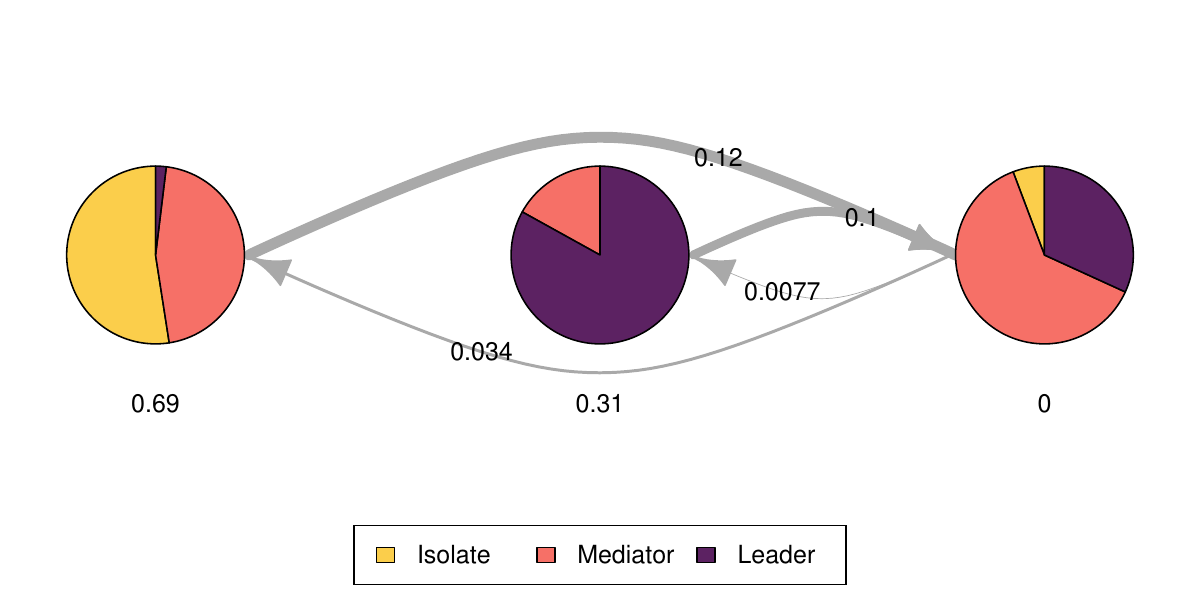}

}

\caption{\label{fig-hmm-2}HMM with three hidden states.}

\end{figure}

Interpreting the results in Figure~\ref{fig-hmm-2} we see that the first
hidden state represents about equal amounts of isolate and mediator
roles, the second hidden state represents mainly Leaders and some
Mediator roles, and the third hidden state represents mainly Mediator
roles and partly Leader roles. Interestingly, none of the students start
as Mediator/Leader, while of the other two the Isolate/Mediator state is
more typical (two thirds). There are no transitions from the first to
the second state nor vice versa, and transition probabilities to the
second state are considerably higher than away from it. In other words,
it seems that the model has two different origin states and one
destination state.

\begin{figure}[H]

{\centering \includegraphics{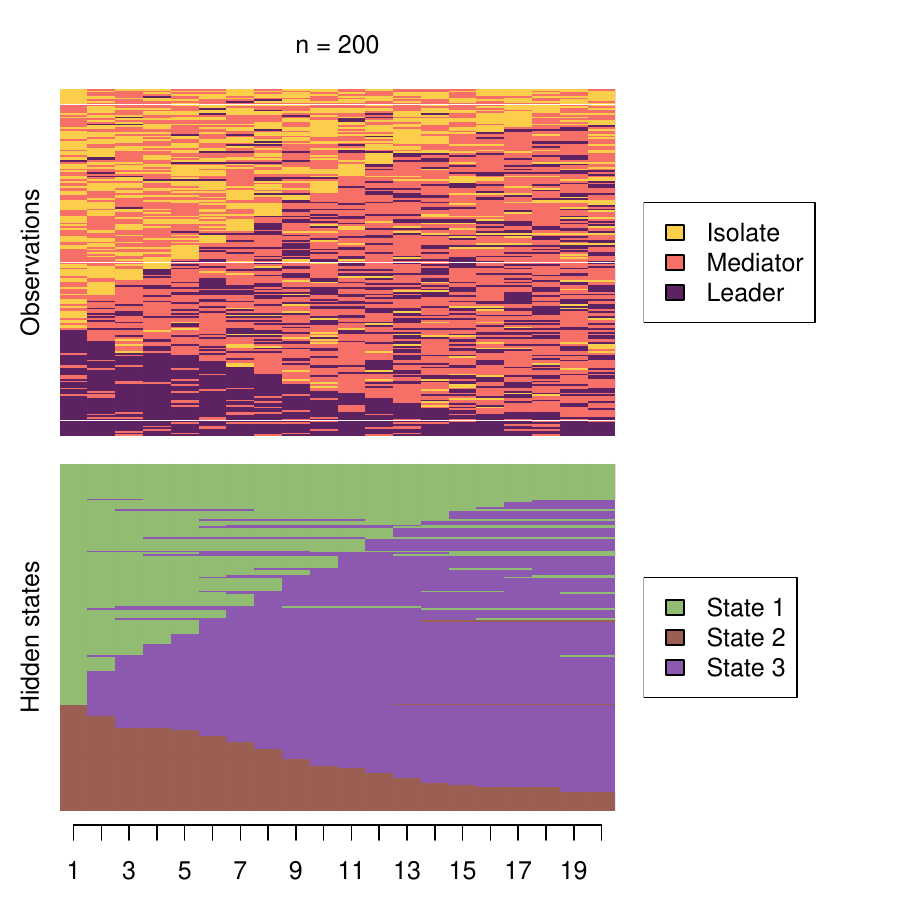}

}

\caption{\label{fig-hmm-ssplot}Observed and hidden state sequences from
the HMM with three hidden states.}

\end{figure}

We can visualise the observed and/or hidden state sequences with the
\texttt{ssplot()} function. The \texttt{ssplot()} function can take an
\texttt{stslist} object or a model object of class \texttt{mm} or
\texttt{hmm} (see Figure~\ref{fig-hmm-ssplot}). Here we want to plot
full sequence index plots (\texttt{type\ =\ "I"}) of both observed and
hidden states (\texttt{plots\ =\ "both"}) and sort the sequences using
multidimensional scaling of hidden states
(\texttt{sortv\ =\ "mds.hidden"}). See the \texttt{seqHMM} manual and
visualisation vignette for more information on the different plotting
options.

\begin{Shaded}
\begin{Highlighting}[]
\FunctionTok{ssplot}\NormalTok{(fit\_hmm\_2}\SpecialCharTok{$}\NormalTok{model, }
       \CommentTok{\# Plot sequence index plot (full sequences)}
       \AttributeTok{type =} \StringTok{"I"}\NormalTok{, }
       \CommentTok{\# Plot observed and hidden state sequences}
       \AttributeTok{plots =} \StringTok{"both"}\NormalTok{, }
       \CommentTok{\# Sort sequences by the scores of multidimensional scaling}
       \AttributeTok{sortv =} \StringTok{"mds.hidden"}\NormalTok{,}
       \CommentTok{\# X axis tick labels}
       \AttributeTok{xtlab =} \DecValTok{1}\SpecialCharTok{:}\DecValTok{20}\NormalTok{)}
\end{Highlighting}
\end{Shaded}

By looking at the sequences, we can see that even though none of the
students start in hidden state 3, the majority of them transition there.
In the end, most students end up alternating between mediating and
leadership roles.

Is the three-state model better than the two-state model? As already
mentioned, we can use model selection criteria to test that. To make
sure that the three-state model is the best, we also estimate a HMM with
four hidden states and then use the Bayesian information criterion for
comparing between the three models. Because the four-state model is more
complex, we increase the number of re-estimation rounds for the EM
algorithm to 100.

\begin{Shaded}
\begin{Highlighting}[]
\CommentTok{\# Set seed for randomisation}
\FunctionTok{set.seed}\NormalTok{(}\DecValTok{1}\NormalTok{)}

\CommentTok{\# Build and estimate a HMM with four states}
\NormalTok{hidden\_markov\_model\_3 }\OtherTok{\textless{}{-}} \FunctionTok{build\_hmm}\NormalTok{(roles\_seq, }\AttributeTok{n\_states =} \DecValTok{4}\NormalTok{)}

\NormalTok{fit\_hmm\_3 }\OtherTok{\textless{}{-}} \FunctionTok{fit\_model}\NormalTok{(hidden\_markov\_model\_3, }
                       \AttributeTok{control\_em =} \FunctionTok{list}\NormalTok{(}\AttributeTok{restart =} \FunctionTok{list}\NormalTok{(}\AttributeTok{times =} \DecValTok{100}\NormalTok{)))}

\NormalTok{fit\_hmm\_3}\SpecialCharTok{$}\NormalTok{em\_results}\SpecialCharTok{$}\NormalTok{best\_opt\_restart}
\end{Highlighting}
\end{Shaded}

\begin{verbatim}
 [1] -3534.304 -3534.304 -3534.304 -3534.304 -3534.304 -3534.304 -3534.304
 [8] -3534.304 -3534.304 -3534.304 -3534.304 -3534.304 -3534.304 -3534.305
[15] -3534.305 -3534.306 -3534.308 -3534.310 -3534.332 -3534.335 -3534.335
[22] -3534.335 -3534.336 -3534.337 -3534.337
\end{verbatim}

The best model was found only 13 times out of 101 estimation rounds from
randomised starting values. A cautious researcher might be wise to opt
for a higher number of estimation rounds for increased certainty, but
here we will proceed to calculating the BIC values.

\begin{Shaded}
\begin{Highlighting}[]
\FunctionTok{BIC}\NormalTok{(fit\_hmm}\SpecialCharTok{$}\NormalTok{model)}
\end{Highlighting}
\end{Shaded}

\begin{verbatim}
[1] 7430.028
\end{verbatim}

\begin{Shaded}
\begin{Highlighting}[]
\FunctionTok{BIC}\NormalTok{(fit\_hmm\_2}\SpecialCharTok{$}\NormalTok{model)}
\end{Highlighting}
\end{Shaded}

\begin{verbatim}
[1] 7208.427
\end{verbatim}

\begin{Shaded}
\begin{Highlighting}[]
\FunctionTok{BIC}\NormalTok{(fit\_hmm\_3}\SpecialCharTok{$}\NormalTok{model)}
\end{Highlighting}
\end{Shaded}

\begin{verbatim}
[1] 7259.37
\end{verbatim}

Generally speaking, the lower the BIC, the better the model. We can see
that the three-state model (\texttt{fit\_hmm\_2}) has the lowest BIC
value, so three clusters is the best choice (at least among HMMs with
2--4 hidden states).

\hypertarget{mixture-markov-models}{%
\subsubsection{Mixture Markov models}\label{mixture-markov-models}}

The MMM can be defined with the \texttt{build\_mmm()} function.
Similarly to HMMs, we need to either give the number of clusters with
the \texttt{n\_clusters} argument, which generates random starting
values for the parameter estimates, or give starting values manually as
\texttt{initial\_probs} and \texttt{transition\_probs}. Here we use
random starting values:

\begin{Shaded}
\begin{Highlighting}[]
\CommentTok{\# Set seed for randomisation}
\FunctionTok{set.seed}\NormalTok{(}\DecValTok{123}\NormalTok{)}
\CommentTok{\# Define model structure (3 clusters)}
\NormalTok{mmm }\OtherTok{\textless{}{-}} \FunctionTok{build\_mmm}\NormalTok{(roles\_seq, }\AttributeTok{n\_clusters =} \DecValTok{3}\NormalTok{)}
\end{Highlighting}
\end{Shaded}

Again, the model is estimated with the \texttt{fit\_model()} function:

\begin{Shaded}
\begin{Highlighting}[]
\NormalTok{fit\_mmm }\OtherTok{\textless{}{-}} \FunctionTok{fit\_model}\NormalTok{(mmm)}
\end{Highlighting}
\end{Shaded}

\begin{figure}[H]

{\centering \includegraphics{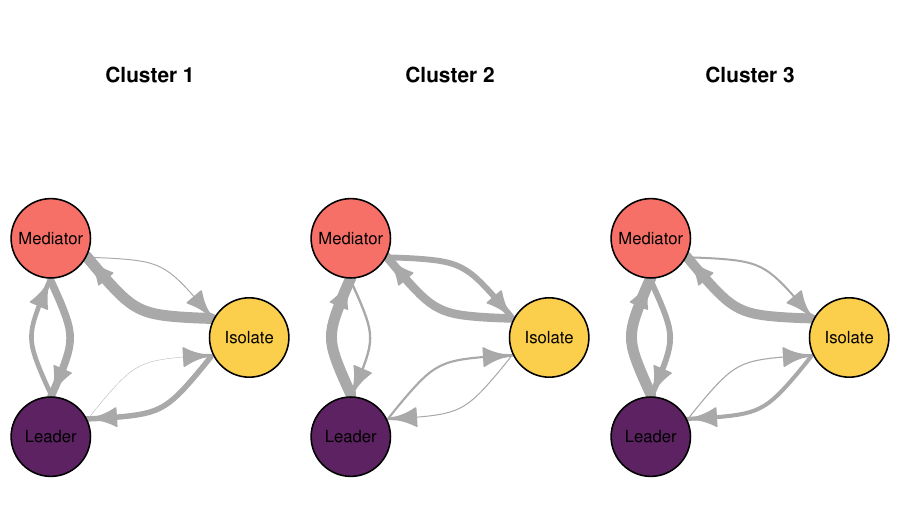}

}

\caption{\label{fig-mmm-pie}MMM with three clusters.}

\end{figure}

The results for each cluster can be plotted one at a time
(interactively, the default), or in one joint figure. Here we opt for
the latter (see Figure~\ref{fig-mmm-pie}). At the same time we also
illustrate some other plotting options:

\begin{Shaded}
\begin{Highlighting}[]
\FunctionTok{plot}\NormalTok{(fit\_mmm}\SpecialCharTok{$}\NormalTok{model, }
     \CommentTok{\# Plot all clusters at the same time}
     \AttributeTok{interactive =} \ConstantTok{FALSE}\NormalTok{, }
     \CommentTok{\# Set the number of rows and columns for cluster plots (one row, three columns)}
     \AttributeTok{nrow =} \DecValTok{1}\NormalTok{, }\AttributeTok{ncol =} \DecValTok{3}\NormalTok{,}
     \CommentTok{\# Omit legends}
     \AttributeTok{with.legend =} \ConstantTok{FALSE}\NormalTok{, }
     \CommentTok{\# Choose another layout for the vertices (see plot.igraph)}
     \AttributeTok{layout =}\NormalTok{ layout\_in\_circle,}
     \CommentTok{\# Omit pie graphs from vertices}
     \AttributeTok{pie =} \ConstantTok{FALSE}\NormalTok{,}
     \CommentTok{\# Set state colours}
     \AttributeTok{vertex.color =} \FunctionTok{cpal}\NormalTok{(roles\_seq),}
     \CommentTok{\# Increase the size of the circle}
     \AttributeTok{vertex.size =} \DecValTok{80}\NormalTok{,}
     \CommentTok{\# Plot state labels instead of initial probabilities}
     \AttributeTok{vertex.label =} \StringTok{"names"}\NormalTok{, }
     \CommentTok{\# Choose font colour for state labels}
     \AttributeTok{vertex.label.color =} \StringTok{"black"}\NormalTok{, }
     \CommentTok{\# Set state label in the centre of the circle}
     \AttributeTok{vertex.label.dist =} \DecValTok{0}\NormalTok{,}
     \CommentTok{\# Omit labels for transition probabilities}
     \AttributeTok{edge.label =} \ConstantTok{NA}\NormalTok{)}
\end{Highlighting}
\end{Shaded}

\begin{figure}

\begin{minipage}[t]{0.50\linewidth}

{\centering 

\raisebox{-\height}{

\includegraphics{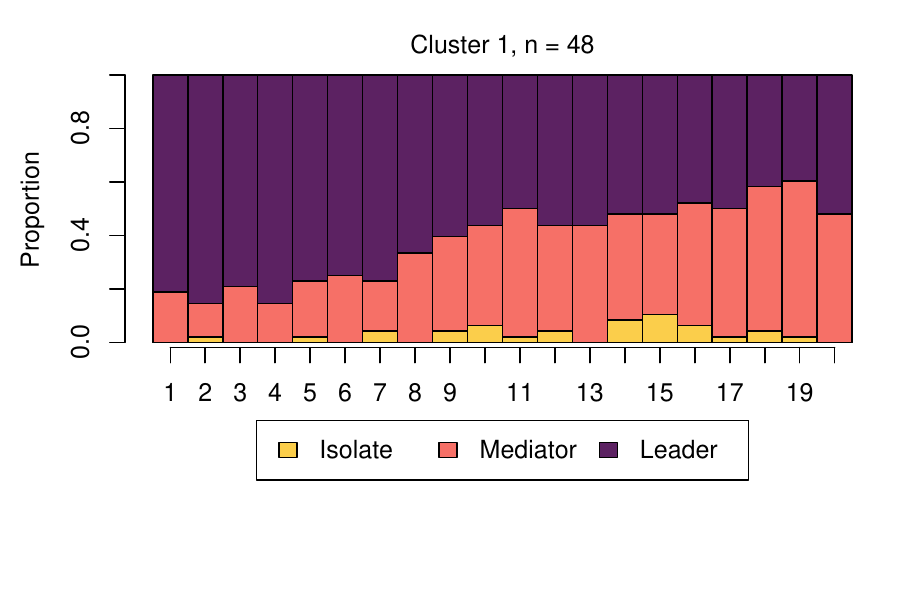}

}

}

\subcaption{\label{fig-mmm-mss-1}Cluster 1.}
\end{minipage}%
\begin{minipage}[t]{0.50\linewidth}

{\centering 

\raisebox{-\height}{

\includegraphics{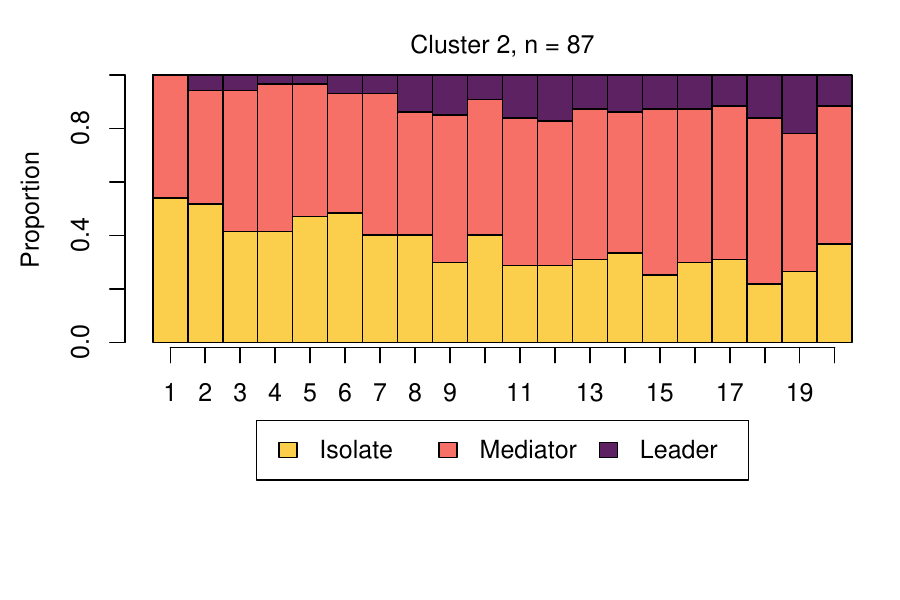}

}

}

\subcaption{\label{fig-mmm-mss-2}Cluster 2.}
\end{minipage}%
\newline
\begin{minipage}[t]{0.50\linewidth}

{\centering 

\raisebox{-\height}{

\includegraphics{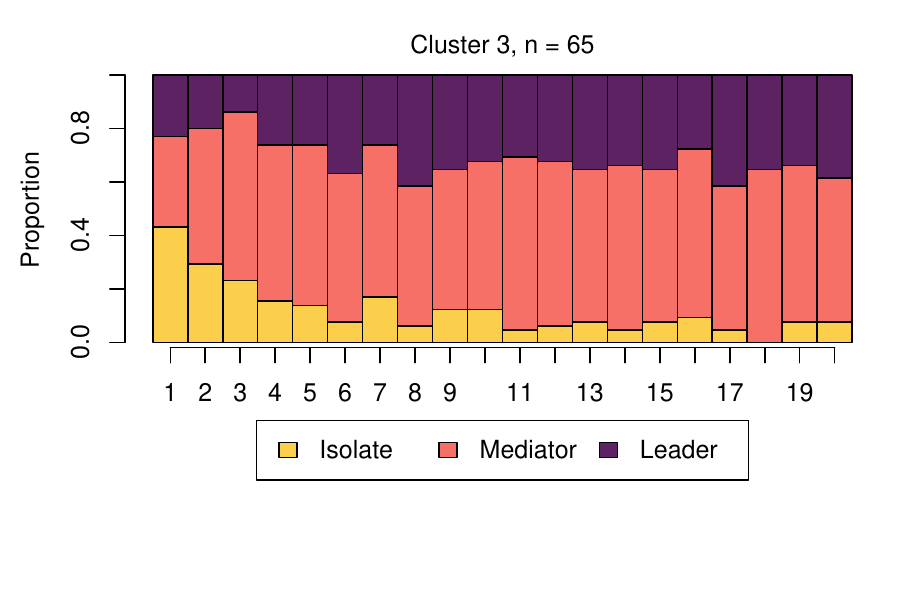}

}

}

\subcaption{\label{fig-mmm-mss-3}Cluster 3.}
\end{minipage}%

\caption{\label{fig-mmm-mss}State distribution plots by most probable
clusters estimated with the mixture Markov model.}

\end{figure}

The following code plots the sequence distribution plot of each cluster
(Figure~\ref{fig-mmm-mss}). In Cluster 1, we see low probabilities to
downward mobility and high probabilities for upward mobility, so this
cluster describes leadership trajectories. In Cluster 2, we can see that
the thickest arrows lead to mediator and isolates roles, so this cluster
describes trajectories with less central roles in collaboration. In
Cluster 3, we see the highest transition probabilities for entering the
mediator role but also some transitions from mediator to leader, so this
cluster describes trajectories with more moderate levels of
participation in comparison to cluster 1. This behavior is easier to see
when visualising the sequences in their most probable clusters. The plot
is interactive, so we need to hit `Enter' on the console to generate
each plot. Alternatively, we can specify which cluster we want to plot
using the \texttt{which.plots} argument.

\begin{Shaded}
\begin{Highlighting}[]
\NormalTok{cl1 }\OtherTok{\textless{}{-}} \FunctionTok{mssplot}\NormalTok{(fit\_mmm}\SpecialCharTok{$}\NormalTok{model, }
               \CommentTok{\# Plot Y axis}
               \AttributeTok{yaxis =} \ConstantTok{TRUE}\NormalTok{, }
               \CommentTok{\# Legend position}
               \AttributeTok{with.legend  =} \StringTok{"bottom"}\NormalTok{,}
               \CommentTok{\# Legend columns}
               \AttributeTok{ncol.legend =} \DecValTok{3}\NormalTok{,}
               \CommentTok{\# Label for Y axis}
               \AttributeTok{ylab =} \StringTok{"Proportion"}\NormalTok{)}
\end{Highlighting}
\end{Shaded}

We can add covariates to the model to explain cluster membership
probabilities. For this, we need to provide a data frame (argument
\texttt{data}) and the corresponding formula (argument
\texttt{formula}). In the example data we use the data frame called
\texttt{cov\_data} that we created at the beginning of the tutorial with
columns \texttt{ID} and \texttt{GPA}, where the order of the \texttt{ID}
variable matches to that of the sequence data \texttt{roles\_seq} (note
that the \texttt{ID} variable is not used in the model building, so the
user needs to make sure that both matrices are sorted by ID). We can now
use the information about students' GPA level as a predictor of the
cluster memberships.

Numerical estimation of complex models from random starting values may
lead to convergence issues and other problems in the estimation (you
may, for example, get warnings about the EM algorithm failing). To avoid
such issues, giving informative starting values is often helpful. This
model is more complex than the model without covariates and estimation
from random starting values leads to convergence issues (not shown
here). To facilitate model estimation, we use the results from the
previous MMM as informative starting values. Here we also remove the
common intercept by adding \texttt{0} to the \texttt{formula}, which
simplifies the interpretation of the covariate effects later (instead of
comparing to a reference category, we get separate coefficients for each
of the three GPA categories).

\begin{Shaded}
\begin{Highlighting}[]
\FunctionTok{set.seed}\NormalTok{(}\DecValTok{98765}\NormalTok{)}
\NormalTok{mmm\_2 }\OtherTok{\textless{}{-}} \FunctionTok{build\_mmm}\NormalTok{(roles\_seq, }
                   \CommentTok{\# Starting values for initial probabilities}
                   \AttributeTok{initial\_probs =}\NormalTok{ fit\_mmm}\SpecialCharTok{$}\NormalTok{model}\SpecialCharTok{$}\NormalTok{initial\_probs,}
                   \CommentTok{\# Starting values for transition probabilities}
                   \AttributeTok{transition\_probs =}\NormalTok{ fit\_mmm}\SpecialCharTok{$}\NormalTok{model}\SpecialCharTok{$}\NormalTok{transition\_probs,}
                   \CommentTok{\# Data frame for covariates}
                   \AttributeTok{data =}\NormalTok{ cov\_data, }
                   \CommentTok{\# Formula for covariates (one{-}sided)}
                   \AttributeTok{formula =} \SpecialCharTok{\textasciitilde{}} \DecValTok{0} \SpecialCharTok{+}\NormalTok{ GPA)}
\end{Highlighting}
\end{Shaded}

Again, the model is estimated with the \texttt{fit\_model()} function.
Here we use the EM algorithm with 50 restarts from random starting
values:

\begin{Shaded}
\begin{Highlighting}[]
\FunctionTok{set.seed}\NormalTok{(}\DecValTok{12345}\NormalTok{)}
\NormalTok{fit\_mmm\_2 }\OtherTok{\textless{}{-}} \FunctionTok{fit\_model}\NormalTok{(mmm\_2, }
                       \CommentTok{\# EM with randomised restarts}
                       \AttributeTok{control\_em =} \FunctionTok{list}\NormalTok{(}\AttributeTok{restart =} \FunctionTok{list}\NormalTok{(}
                       \CommentTok{\# 50 restarts}
                       \AttributeTok{times =} \DecValTok{50}\NormalTok{, }
                       \CommentTok{\# Store loglik values from all 50 + 1 estimation rounds}
                       \AttributeTok{n\_optimum =} \DecValTok{51}\NormalTok{)}
\NormalTok{                      ))}
\end{Highlighting}
\end{Shaded}

\begin{verbatim}
Warning in fit_model(mmm_2, control_em = list(restart = list(times = 50, : EM
algorithm failed: Estimation of gamma coefficients failed due to singular
Hessian.
\end{verbatim}

The model was estimated 50 + 1 times (first from the starting values we
provided and then from 50 randomised values). We get one warning about
the EM algorithm failing. However, 50 estimation rounds were successful.
We can check that the best model was found several times from different
starting values (37 times, to be precise):

\begin{Shaded}
\begin{Highlighting}[]
\NormalTok{fit\_mmm\_2}\SpecialCharTok{$}\NormalTok{em\_results}\SpecialCharTok{$}\NormalTok{best\_opt\_restart}
\end{Highlighting}
\end{Shaded}

\begin{verbatim}
 [1] -3614.627 -3614.627 -3614.627 -3614.627 -3614.627 -3614.627 -3614.627
 [8] -3614.627 -3614.627 -3614.627 -3614.627 -3614.627 -3614.627 -3614.627
[15] -3614.627 -3614.627 -3614.627 -3614.627 -3614.627 -3614.627 -3614.627
[22] -3614.627 -3614.627 -3614.627 -3614.627 -3614.627 -3614.627 -3614.627
[29] -3614.627 -3614.627 -3614.627 -3614.627 -3614.627 -3614.627 -3614.627
[36] -3614.627 -3614.627 -3619.695 -3624.547 -3624.547 -3624.547 -3624.547
[43] -3624.547 -3624.547 -3624.547 -3624.547 -3624.547 -3624.547 -3631.328
[50] -3637.344      -Inf
\end{verbatim}

\begin{figure}

\begin{minipage}[t]{0.50\linewidth}

{\centering 

\raisebox{-\height}{

\includegraphics{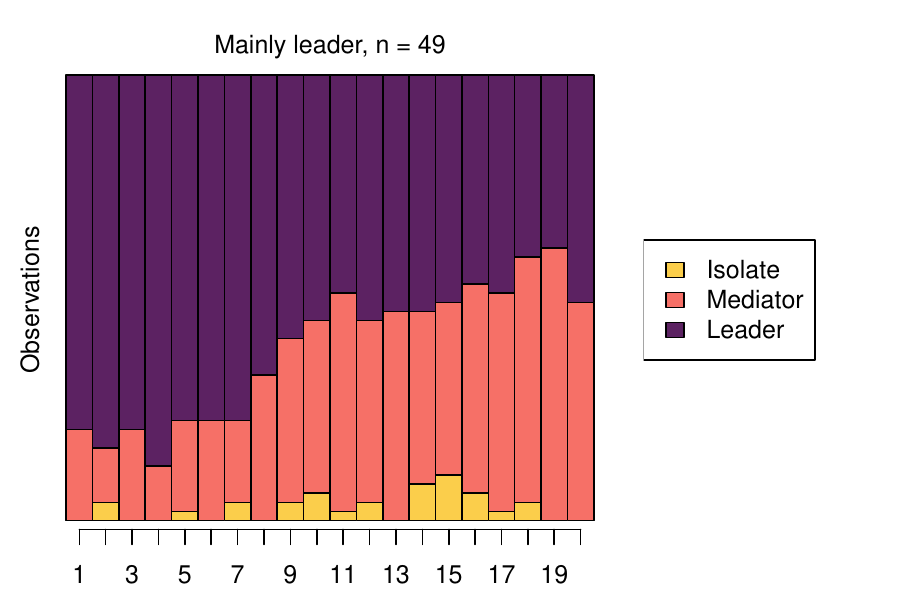}

}

}

\subcaption{\label{fig-mmm2-mss-1}Mainly leader.}
\end{minipage}%
\begin{minipage}[t]{0.50\linewidth}

{\centering 

\raisebox{-\height}{

\includegraphics{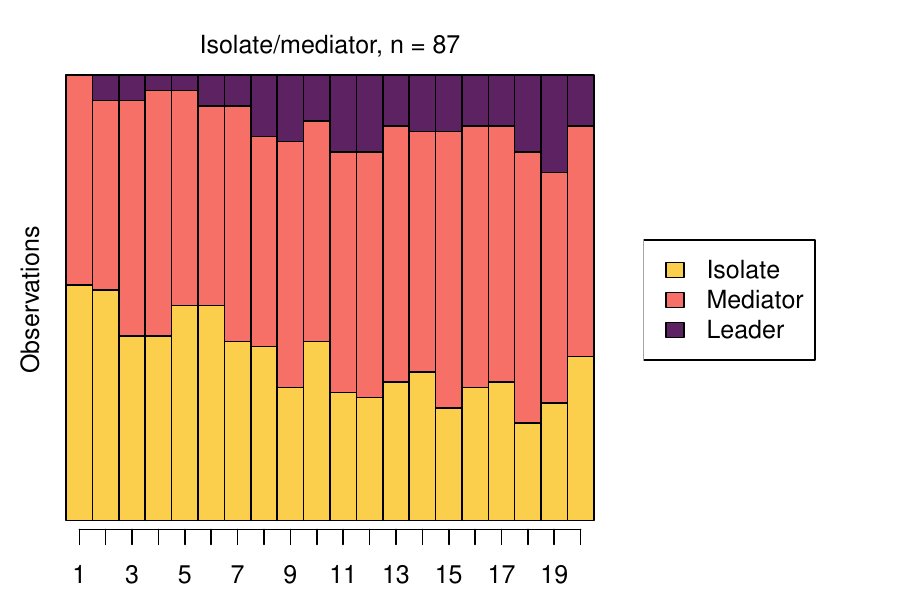}

}

}

\subcaption{\label{fig-mmm2-mss-2}Isolate/mediator.}
\end{minipage}%
\newline
\begin{minipage}[t]{0.50\linewidth}

{\centering 

\raisebox{-\height}{

\includegraphics{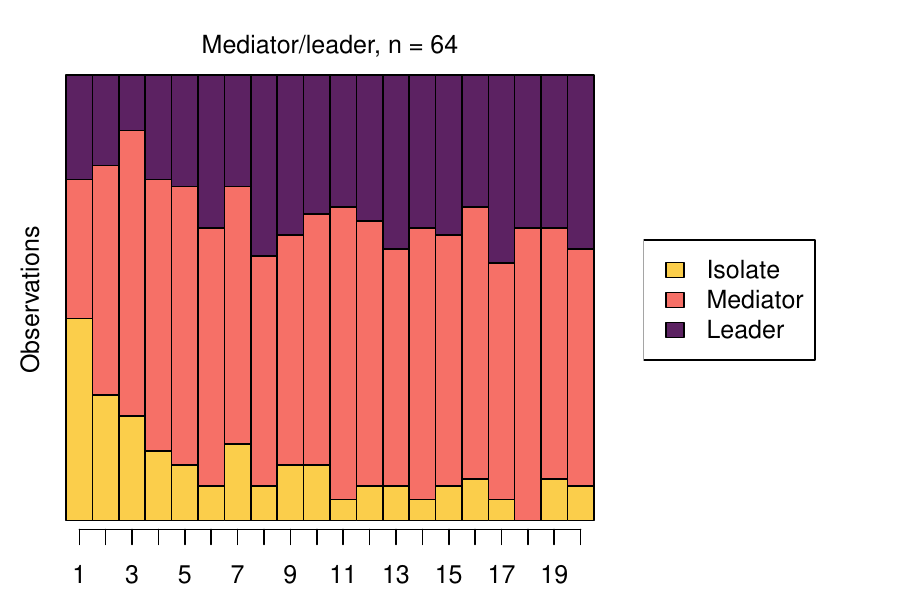}

}

}

\subcaption{\label{fig-mmm2-mss-3}Mediator/leader.}
\end{minipage}%

\caption{\label{fig-mmm2-mss}State distribution plots by most probable
clusters estimated with the mixture Markov model with covariates.}

\end{figure}

We can now be fairly certain that the optimal model has been found, and
can proceed to interpreting the results. The clusters are very similar
to what we found before. We can give the clusters more informative
labels and then show state distribution plots in each cluster in
Figure~\ref{fig-mmm2-mss}:

\begin{Shaded}
\begin{Highlighting}[]
\FunctionTok{cluster\_names}\NormalTok{(fit\_mmm\_2}\SpecialCharTok{$}\NormalTok{model) }\OtherTok{\textless{}{-}} \FunctionTok{c}\NormalTok{(}\StringTok{"Mainly leader"}\NormalTok{, }\StringTok{"Isolate/mediator"}\NormalTok{, }\StringTok{"Mediator/leader"}\NormalTok{)}
\FunctionTok{mssplot}\NormalTok{(fit\_mmm\_2}\SpecialCharTok{$}\NormalTok{model)}
\end{Highlighting}
\end{Shaded}

The model summary shows information about parameter estimates of
covariates and prior and posterior cluster membership probabilities
(these refer to cluster membership probabilities before or after
conditioning on the observed sequences, respectively):

\begin{Shaded}
\begin{Highlighting}[]
\NormalTok{summary\_mmm\_2 }\OtherTok{\textless{}{-}} \FunctionTok{summary}\NormalTok{(fit\_mmm\_2}\SpecialCharTok{$}\NormalTok{model)}
\NormalTok{summary\_mmm\_2}
\end{Highlighting}
\end{Shaded}

\begin{verbatim}
Covariate effects :
Mainly leader is the reference.

Isolate/mediator :
           Estimate  Std. error
GPALow       1.9221       0.478
GPAMiddle    0.3901       0.314
GPAHigh     -0.0451       0.277

Mediator/leader :
           Estimate  Std. error
GPALow        1.670       0.487
GPAMiddle     0.411       0.312
GPAHigh      -0.667       0.332

Log-likelihood: -3614.627   BIC: 7461.487 

Means of prior cluster probabilities :
   Mainly leader Isolate/mediator  Mediator/leader 
           0.244            0.425            0.331 

Most probable clusters :
            Mainly leader  Isolate/mediator  Mediator/leader
count                  49                87               64
proportion          0.245             0.435             0.32

Classification table :
Mean cluster probabilities (in columns) by the most probable cluster (rows)

                 Mainly leader Isolate/mediator Mediator/leader
Mainly leader          0.91758          0.00136          0.0811
Isolate/mediator       0.00081          0.89841          0.1008
Mediator/leader        0.05902          0.10676          0.8342
\end{verbatim}

We will first interpret the information on prior and posterior cluster
membership probabilities and then proceed to interpreting covariate
effects. Firstly, the \texttt{means\ of\ prior\ cluster\ probabilities}
give information on how likely each cluster is in the whole population
of students (33\% in Mediator, 24\% in Leader, and 43\% in Isolate).
Secondly, \texttt{Most\ probable\ clusters} shows group sizes and
proportions if each student would be classified into the cluster for
which they have the highest cluster membership probability.

Thirdly, the \texttt{Classification\ table} shows mean cluster
probabilities (in columns) by the most probable cluster (in rows). We
can see that the clusters are fairly crisp (the certainty of cluster
memberships are fairly high) because the membership probabilities are
large in the diagonal of the table. The uncertainty of the
classification is the highest for the Mediator/leader cluster (among
those that had the highest membership probability in that cluster,
average cluster memberships were 84\% for the Mediator/leader cluster,
6\% for the Mainly leader cluster, and 10\% for the Isolate/mediator
cluster) and the highest in the Mainly leader cluster (92\% for the
Mainly leader cluster, 8\% for the Mediator/leader cluster, and 0.1\%
for the Isolate/mediator cluster).

The part titled \texttt{Covariate\ effects} shows the parameter
estimates for the covariates. Interpretation of the values is similar to
that of multinomial logistic regression, meaning that we can interpret
the direction and uncertainty of the effect --relative to the reference
cluster Mainly leader-- but we cannot directly interpret the magnitude
of the effects (the magnitudes are on log-odds scale). We can see that
individuals with low GPA more often end up in the Isolate/mediator
cluster and the Mediator/leader cluster in comparison to the Mainly
leader cluster (i.e., the standard errors are small in comparison to the
parameter estimates), while individuals with high GPA levels end up in
the Mediator/leader cluster less often but are not more or less likely
to end up in the Isolate/mediator cluster. For categorical covariates
such as our \texttt{GPA} variable, we can also easily compute the prior
cluster membership probabilities from the estimates with the following
call:

\begin{Shaded}
\begin{Highlighting}[]
\FunctionTok{exp}\NormalTok{(fit\_mmm\_2}\SpecialCharTok{$}\NormalTok{model}\SpecialCharTok{$}\NormalTok{coefficients)}\SpecialCharTok{/}\FunctionTok{rowSums}\NormalTok{(}\FunctionTok{exp}\NormalTok{(fit\_mmm\_2}\SpecialCharTok{$}\NormalTok{model}\SpecialCharTok{$}\NormalTok{coefficients))}
\end{Highlighting}
\end{Shaded}

\begin{verbatim}
          Mainly leader Isolate/mediator Mediator/leader
GPALow       0.07605453        0.5198587       0.4040868
GPAMiddle    0.25090105        0.3705958       0.3785031
GPAHigh      0.40497185        0.3870997       0.2079285
\end{verbatim}

The matrix shows the levels of the covariates in the rows and the
clusters in the columns. Among the high-GPA students, 41 percent are
classified as Mainly leaders, 39 percent as Isolate/mediators, and 21
percent as Mediator/leaders. Among middle-GPA students classification is
relatively uniform (25\% as Mainly leaders, 37\% as Isolate/mediators
and 38 Mediator/leaders) whereas most of the low-GPA students are
classified as Isolate/mediators or Mediator/leaders (52\% and 40\%,
respectively).

The summary object also calculates prior and posterior cluster
memberships for each student. We omit them here, for brevity, but
demonstrate that they can be obtained as follows:

\begin{Shaded}
\begin{Highlighting}[]
\NormalTok{prior\_prob }\OtherTok{\textless{}{-}}\NormalTok{ summary\_mmm\_2}\SpecialCharTok{$}\NormalTok{prior\_cluster\_probabilities}
\NormalTok{posterior\_prob }\OtherTok{\textless{}{-}}\NormalTok{ summary\_mmm\_2}\SpecialCharTok{$}\NormalTok{posterior\_cluster\_probabilities}
\end{Highlighting}
\end{Shaded}

\hypertarget{mixture-hidden-markov-models-1}{%
\subsubsection{Mixture hidden Markov
models}\label{mixture-hidden-markov-models-1}}

Finally, we will proceed to the most complex of the models, the MHMM.

For defining a MHMM, we use the \texttt{build\_mhmm()} function. Again,
we can use the argument \texttt{n\_states} which is now a vector showing
the number of hidden states in each cluster (the length of the vector
defines the number of clusters). We will begin by estimating a MHMM with
three clusters, each with two hidden states:

\begin{Shaded}
\begin{Highlighting}[]
\FunctionTok{set.seed}\NormalTok{(}\DecValTok{123}\NormalTok{)}
\NormalTok{mhmm }\OtherTok{\textless{}{-}} \FunctionTok{build\_mhmm}\NormalTok{(roles\_seq, }
                   \AttributeTok{n\_states =} \FunctionTok{c}\NormalTok{(}\DecValTok{2}\NormalTok{, }\DecValTok{2}\NormalTok{, }\DecValTok{2}\NormalTok{),}
                   \AttributeTok{data =}\NormalTok{ cov\_data, }
                   \AttributeTok{formula =} \SpecialCharTok{\textasciitilde{}} \DecValTok{0} \SpecialCharTok{+}\NormalTok{ GPA)}
\NormalTok{fit\_mhmm }\OtherTok{\textless{}{-}} \FunctionTok{fit\_model}\NormalTok{(mhmm)}
\end{Highlighting}
\end{Shaded}

\begin{verbatim}
Error in fit_model(mhmm): EM algorithm failed: Estimation of gamma coefficients failed due to singular Hessian.
\end{verbatim}

In this case, we get an error message about the EM algorithm failing.
This means that the algorithm was not able to find parameter estimates
from the random starting values the \texttt{build\_mhmm()} function
generated and we need to adjust our code.

Starting values for the parameters of the MHMM can be given with the the
arguments \texttt{initial\_probs}, \texttt{transition\_probs}, and
\texttt{emission\_probs}. For the MHMM, these are lists of vectors and
matrices, one for each cluster. We use the same number of hidden states
(two) for each cluster. We define the initial values for the transition
and emission probabilities as well as regression coefficients ourselves.
We also restrict the initial state probabilities so that in each cluster
every student is forced to start from the same (first) hidden state.

\begin{Shaded}
\begin{Highlighting}[]
\FunctionTok{set.seed}\NormalTok{(}\DecValTok{1}\NormalTok{)}

\CommentTok{\# Set initial probabilities}
\NormalTok{init }\OtherTok{\textless{}{-}} \FunctionTok{list}\NormalTok{(}\FunctionTok{c}\NormalTok{(}\DecValTok{1}\NormalTok{, }\DecValTok{0}\NormalTok{), }\FunctionTok{c}\NormalTok{(}\DecValTok{1}\NormalTok{, }\DecValTok{0}\NormalTok{), }\FunctionTok{c}\NormalTok{(}\DecValTok{1}\NormalTok{, }\DecValTok{0}\NormalTok{))}

\CommentTok{\# Define own transition probabilities}
\NormalTok{trans }\OtherTok{\textless{}{-}} \FunctionTok{matrix}\NormalTok{(}\FunctionTok{c}\NormalTok{(}
  \FloatTok{0.9}\NormalTok{, }\FloatTok{0.1}\NormalTok{,}
  \FloatTok{0.1}\NormalTok{, }\FloatTok{0.9}
\NormalTok{), }\AttributeTok{nrow =} \DecValTok{2}\NormalTok{, }\AttributeTok{byrow =} \ConstantTok{TRUE}\NormalTok{)}

\NormalTok{translist }\OtherTok{\textless{}{-}} \FunctionTok{list}\NormalTok{(trans, trans, trans)}

\CommentTok{\# Simulate emission probabilities}
\NormalTok{emiss }\OtherTok{\textless{}{-}} \FunctionTok{simulate\_emission\_probs}\NormalTok{(}\AttributeTok{n\_states =} \FunctionTok{c}\NormalTok{(}\DecValTok{2}\NormalTok{, }\DecValTok{2}\NormalTok{, }\DecValTok{2}\NormalTok{), }
                                 \AttributeTok{n\_symbols =} \DecValTok{3}\NormalTok{, }
                                 \AttributeTok{n\_clusters =} \DecValTok{3}\NormalTok{)}

\NormalTok{emiss }\OtherTok{\textless{}{-}} \FunctionTok{replicate}\NormalTok{(}\DecValTok{3}\NormalTok{, }\FunctionTok{matrix}\NormalTok{(}\DecValTok{1}\SpecialCharTok{/}\DecValTok{3}\NormalTok{, }\DecValTok{2}\NormalTok{, }\DecValTok{3}\NormalTok{), }\AttributeTok{simplify =} \ConstantTok{FALSE}\NormalTok{)}

\CommentTok{\# Define initial values for coefficients}
\CommentTok{\# Here we start from a case where low GPA correlates with Cluster 1, }
\CommentTok{\# whereas middle and high GPA has no effect}
\NormalTok{beta }\OtherTok{\textless{}{-}} \FunctionTok{cbind}\NormalTok{(}\DecValTok{0}\NormalTok{, }\FunctionTok{c}\NormalTok{(}\SpecialCharTok{{-}}\DecValTok{2}\NormalTok{, }\DecValTok{0}\NormalTok{, }\DecValTok{0}\NormalTok{), }\FunctionTok{c}\NormalTok{(}\SpecialCharTok{{-}}\DecValTok{2}\NormalTok{, }\DecValTok{0}\NormalTok{, }\DecValTok{0}\NormalTok{))}

\CommentTok{\# Define model structure}
\NormalTok{mhmm\_2 }\OtherTok{\textless{}{-}} \FunctionTok{build\_mhmm}\NormalTok{(roles\_seq, }
                     \AttributeTok{initial\_probs =}\NormalTok{ init, }\AttributeTok{transition\_probs =}\NormalTok{ translist, }
                     \AttributeTok{emission\_probs =}\NormalTok{ emiss, }\AttributeTok{data =}\NormalTok{ cov\_data, }
                     \AttributeTok{formula =} \SpecialCharTok{\textasciitilde{}} \DecValTok{0} \SpecialCharTok{+}\NormalTok{ GPA, }\AttributeTok{beta =}\NormalTok{ beta)}
\end{Highlighting}
\end{Shaded}

Now that we have built the MHMM, we can estimate its parameters:

\begin{Shaded}
\begin{Highlighting}[]
\FunctionTok{set.seed}\NormalTok{(}\DecValTok{1}\NormalTok{)}
\FunctionTok{suppressWarnings}\NormalTok{(fit\_mhmm\_2 }\OtherTok{\textless{}{-}} \FunctionTok{fit\_model}\NormalTok{(}
\NormalTok{  mhmm\_2,}
  \AttributeTok{control\_em =} \FunctionTok{list}\NormalTok{(}\AttributeTok{restart =} \FunctionTok{list}\NormalTok{(}\AttributeTok{times =} \DecValTok{100}\NormalTok{, }\AttributeTok{n\_optimum =} \DecValTok{101}\NormalTok{)))}
\NormalTok{)}
\end{Highlighting}
\end{Shaded}

We can now check how many times the log-likelihood values occurred in
the 101 estimations:

\begin{Shaded}
\begin{Highlighting}[]
\FunctionTok{table}\NormalTok{(}\FunctionTok{round}\NormalTok{(fit\_mhmm\_2}\SpecialCharTok{$}\NormalTok{em\_results}\SpecialCharTok{$}\NormalTok{best\_opt\_restart, }\DecValTok{2}\NormalTok{))}
\end{Highlighting}
\end{Shaded}

\begin{verbatim}

    -Inf -3672.25 -3595.82 -3588.58 -3584.14 -3526.42 -3525.06 -3519.53 
      56        2        1        3        1        4        1        2 
 -3519.5 -3519.24 
      15       16 
\end{verbatim}

The best model was found 16 times out of 101 times, although the second
beset model with log-likelihood of -3519.5 is likely almost
indistinguishable from the optimal model (-3519.24) as their
log-likelihoods are so close to each other.

\begin{figure}

\begin{minipage}[t]{0.33\linewidth}

{\centering 

\raisebox{-\height}{

\includegraphics{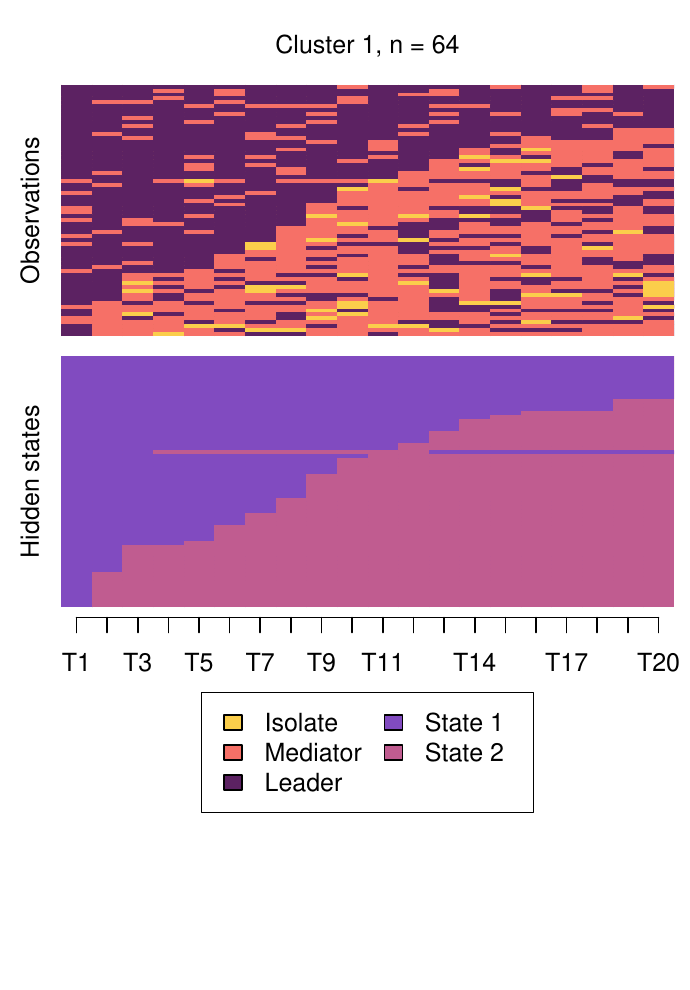}

}

}

\subcaption{\label{fig-mhmm-seq-1}Cluster 1}
\end{minipage}%
\begin{minipage}[t]{0.33\linewidth}

{\centering 

\raisebox{-\height}{

\includegraphics{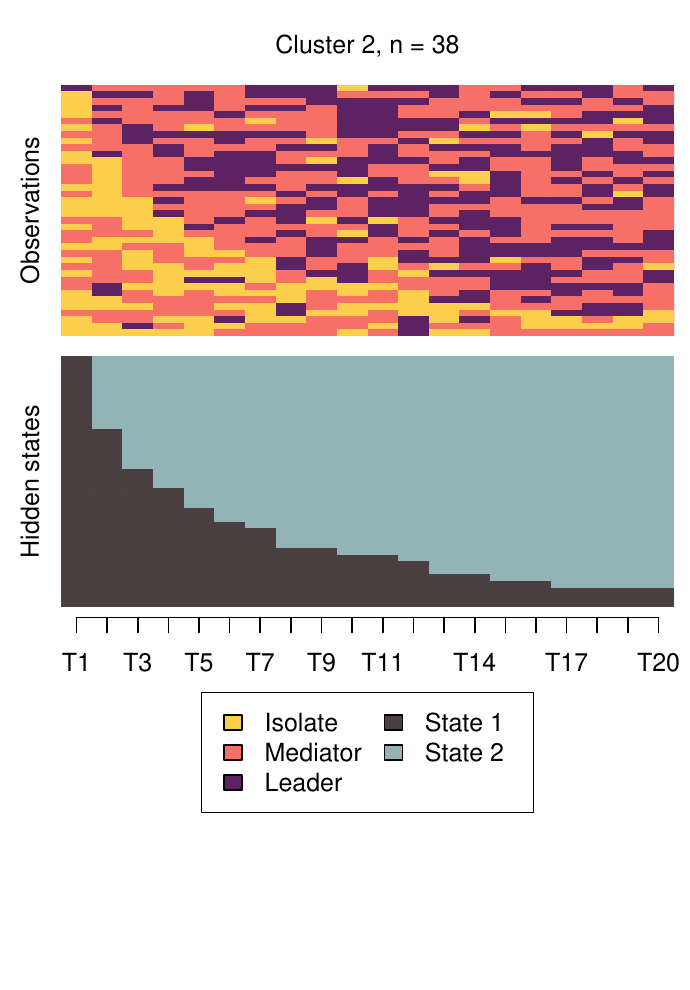}

}

}

\subcaption{\label{fig-mhmm-seq-2}Cluster 2}
\end{minipage}%
\begin{minipage}[t]{0.33\linewidth}

{\centering 

\raisebox{-\height}{

\includegraphics{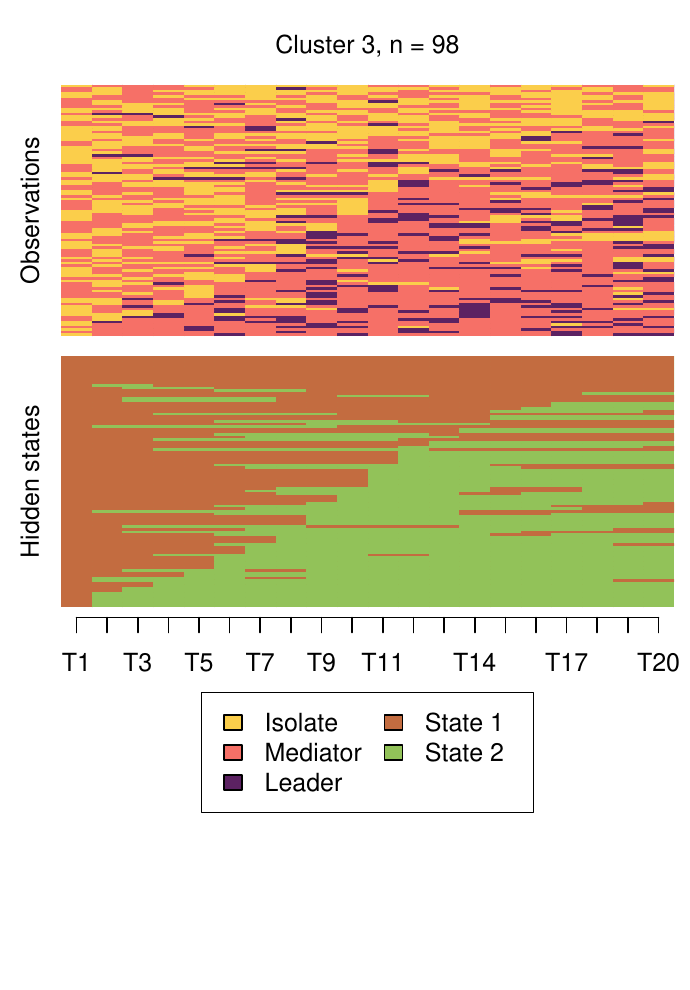}

}

}

\subcaption{\label{fig-mhmm-seq-3}Cluster 3}
\end{minipage}%

\caption{\label{fig-mhmm-seq}MHMM estimated sequence distribution plot
with hidden states.}

\end{figure}

We will start to interpret the model by looking at the sequence plots in
each cluster (see Figure~\ref{fig-mhmm-seq}). The function call is
interactive. As before, if you only want to plot one cluster you can use
the \texttt{which.plots} argument:

\begin{Shaded}
\begin{Highlighting}[]
\FunctionTok{mssplot}\NormalTok{(fit\_mhmm\_2}\SpecialCharTok{$}\NormalTok{model, }\AttributeTok{plots =} \StringTok{"both"}\NormalTok{, }\AttributeTok{type =} \StringTok{"I"}\NormalTok{, }
        \AttributeTok{sortv =} \StringTok{"mds.hidden"}\NormalTok{, }\AttributeTok{with.legend =} \StringTok{"bottom.combined"}\NormalTok{)}
\end{Highlighting}
\end{Shaded}

\begin{figure}[H]

{\centering \includegraphics{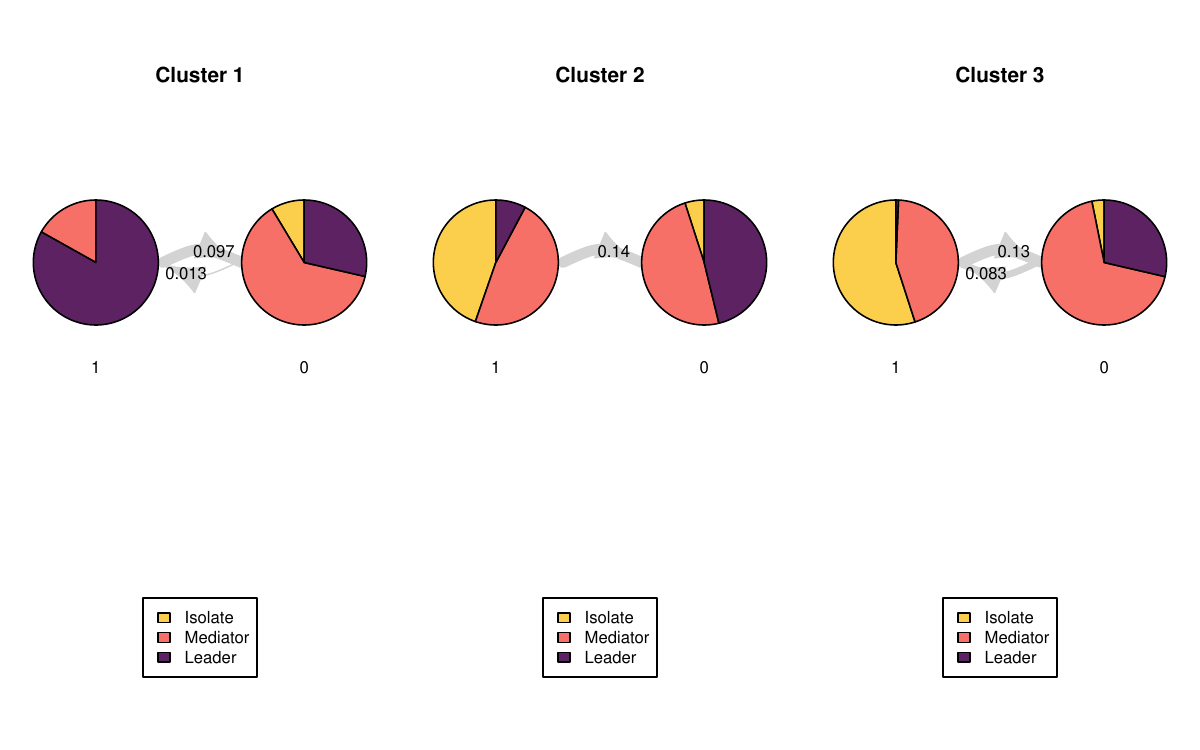}

}

\caption{\label{fig-mhmm-pie}Transitions between states for each
trajectory.}

\end{figure}

We can also visualise the model parameters in each cluster (see
Figure~\ref{fig-mhmm-pie}):

\begin{Shaded}
\begin{Highlighting}[]
\FunctionTok{plot}\NormalTok{(fit\_mhmm\_2}\SpecialCharTok{$}\NormalTok{model, }
     \AttributeTok{vertex.size =} \DecValTok{60}\NormalTok{,}
     \AttributeTok{label.color =} \StringTok{"black"}\NormalTok{,}
     \AttributeTok{vertex.label.color =} \StringTok{"black"}\NormalTok{,}
     \AttributeTok{edge.color =} \StringTok{"lightgray"}\NormalTok{,}
     \AttributeTok{edge.label.color =} \StringTok{"black"}\NormalTok{,}
     \AttributeTok{ncol.legend =} \DecValTok{1}\NormalTok{, }
     \AttributeTok{ncol =} \DecValTok{3}\NormalTok{,}
     \AttributeTok{rescale =} \ConstantTok{FALSE}\NormalTok{,}
     \AttributeTok{interactive =} \ConstantTok{FALSE}\NormalTok{,}
     \AttributeTok{combine.slices =} \DecValTok{0}\NormalTok{)}
\end{Highlighting}
\end{Shaded}

Based on the two plots, we can determine that Cluster 1 describes
students who start as leaders but then transition to alternating between
mediator and leader. Cluster 2 describes students who start by
alternating between isolate and mediator roles and then mainly
transition to alternating between mediator and leader roles. Cluster 3
describes students who start as alternating between isolate and mediator
roles, after which they transition between isolate/mediator and
mediator/leader.

\begin{Shaded}
\begin{Highlighting}[]
\FunctionTok{cluster\_names}\NormalTok{(fit\_mhmm\_2}\SpecialCharTok{$}\NormalTok{model) }\OtherTok{\textless{}{-}} \FunctionTok{c}\NormalTok{(}\StringTok{"Downward transition"}\NormalTok{,}
                                    \StringTok{"Upward transition"}\NormalTok{, }
                                    \StringTok{"Alternating"}\NormalTok{)}
\end{Highlighting}
\end{Shaded}

With \texttt{summary(fit\_mhmm\_2\$model)} we get the parameter
estimates and standard errors for the covariates and information about
clustering:

\begin{Shaded}
\begin{Highlighting}[]
\FunctionTok{summary}\NormalTok{(fit\_mhmm\_2}\SpecialCharTok{$}\NormalTok{model)}
\end{Highlighting}
\end{Shaded}

\begin{verbatim}
Covariate effects :
Downward transition is the reference.

Upward transition :
           Estimate  Std. error
GPALow       -0.455       0.464
GPAMiddle     0.440       0.310
GPAHigh      -2.743       0.727

Alternating :
           Estimate  Std. error
GPALow       1.3560       0.324
GPAMiddle    0.3461       0.316
GPAHigh      0.0468       0.250

Log-likelihood: -3519.243   BIC: 7237.543 

Means of prior cluster probabilities :
Downward transition   Upward transition         Alternating 
              0.302               0.181               0.517 

Most probable clusters :
            Downward transition  Upward transition  Alternating
count                        61                 30          109
proportion                0.305               0.15        0.545

Classification table :
Mean cluster probabilities (in columns) by the most probable cluster (rows)

                    Downward transition Upward transition Alternating
Downward transition             0.95727            0.0267      0.0161
Upward transition               0.03007            0.8037      0.1662
Alternating                     0.00975            0.0962      0.8940
\end{verbatim}

We can see, that the prior probabilities of belonging to each cluster
are very different: half of the students can be described as
alternating, while of the rest, a downward transition is more typical
(31\%). Based on the classification table, the Downward transition
cluster is rather crisp, while the other two are partly overlapping (see
the MMM example for more information on interpreting the classification
table).

The \texttt{Covariate\ effects} tables show that, in comparison to
Alternating cluster, students with low GPA are less likely to end up in
the Upward or Downward transition clusters and students with high GPA
are less likely to end up in Upward transition cluster. Again, we can
calculate the probabilities of belonging to each cluster by GPA levels:

\begin{Shaded}
\begin{Highlighting}[]
\FunctionTok{exp}\NormalTok{(fit\_mhmm\_2}\SpecialCharTok{$}\NormalTok{model}\SpecialCharTok{$}\NormalTok{coefficients)}\SpecialCharTok{/}\FunctionTok{rowSums}\NormalTok{(}\FunctionTok{exp}\NormalTok{(fit\_mhmm\_2}\SpecialCharTok{$}\NormalTok{model}\SpecialCharTok{$}\NormalTok{coefficients))}
\end{Highlighting}
\end{Shaded}

\begin{verbatim}
          Downward transition Upward transition Alternating
GPALow              0.1813217        0.11502283   0.7036555
GPAMiddle           0.2521406        0.39144399   0.3564154
GPAHigh             0.4734128        0.03048189   0.4961054
\end{verbatim}

The table shows that students with low GPA typically belong to the
Alternating cluster (70 \% probability) while students with high GPA
mainly end up in the Downward transition cluster (47\%) or the
Alternating cluster (50\%). Most students with middle GPA end up in the
Upward transition cluster (39\%), but the probabilities are almost as
high for the Alternating cluster (36\%) and also fairly high for the
Downward transition cluster (25\%).

In light of this, it is worth noting that the covariates do not merely
explain the uncovered clusters; as part of the model, they drive the
formation of the clusters. In other words, an otherwise identical model
without the dependence on the \texttt{GPA} covariate may uncover
different groupings with different probabilities.

If we are not sure how many clusters or hidden states we expect, or if
we wish to investigate different combinations of covariates, we can
estimate several models and compare the results with information
criteria or cross-validation. Estimating a large number of complex
models is, however, very time-consuming. Using prior information for
restricting the pool of potential models is useful, and sequence
analysis can also be used as a helpful first step {[}10, 37{]}.

\hypertarget{process}{%
\subsection{Stochastic process mining with Markovian
models}\label{process}}

Process mining can be performed using different methods, techniques and
algorithms. Yet, MMs offer a very powerful framework for process mining
with several advantages over the commonly used methods. First, it is
more theoretically aligned with the idea of a transition from an action
to an action and that actions are temporally dependent on each other.
Second, MMs allow for data to be clustered into similar transition
patterns, a possibility not offered by other process mining methods (see
the process mining chapter of this book). Third, contrary to other
process mining methods, MMs do not require researchers to exclude ---or
trim--- a large part of the data to ``simplify'' the model. For
instance, most of the process mining analyses require an arbitrary
cutoff to trim some transitions so that the process model is readable.
Most importantly, MMs have several fit statistics that we can use to
compare and judge the model fit as we have seen before.

Several R packages can perform stochastic process mining; in this
tutorial we will rely on the same package we discussed earlier and
combine it with a powerful visualization that allows us to effectively
visualize complex processes. In the next example, we will analyse data
extracted from the learning management system logs and offer a detailed
guide to process mining. We will also use MMMs to cluster the data into
latent patterns of transitions. Given that the traditional plotting
function in \texttt{seqHMM} works well with a relatively short alphabet,
we will use a new R package called \texttt{qgraph} for plotting. The
package \texttt{qgraph} offers powerful visualizations which makes
plotting easier, and more interpretable especially for larger models.
Furthermore, \texttt{qgraph} allows researchers to use a fixed layout
for all the plotted networks so the nodes can be compared to each other
more easily.

Let us now go through the analysis. The next chunk of code imports the
prepared sequence data from the sequence analysis chapter. The data
belong to a learning analytics course and the events are coded trace
logs of students' actions such as \emph{Course view},
\emph{Instructions}, \emph{Practicals}, \emph{Social}, etc. Then, we
build a sequence object using the function \texttt{seqdef()} from
\texttt{TraMineR}.

\begin{Shaded}
\begin{Highlighting}[]
\NormalTok{seq\_data }\OtherTok{\textless{}{-}} \FunctionTok{import}\NormalTok{(}\FunctionTok{paste0}\NormalTok{(URL, }\StringTok{"1\_moodleLAcourse/LMS\_data\_wide.xlsx"}\NormalTok{))}
\NormalTok{seq\_data\_all }\OtherTok{\textless{}{-}} \FunctionTok{seqdef}\NormalTok{(seq\_data, }\AttributeTok{var =} \DecValTok{7}\SpecialCharTok{:}\DecValTok{54}\NormalTok{ )}
\end{Highlighting}
\end{Shaded}

\begin{figure}[H]

{\centering \includegraphics{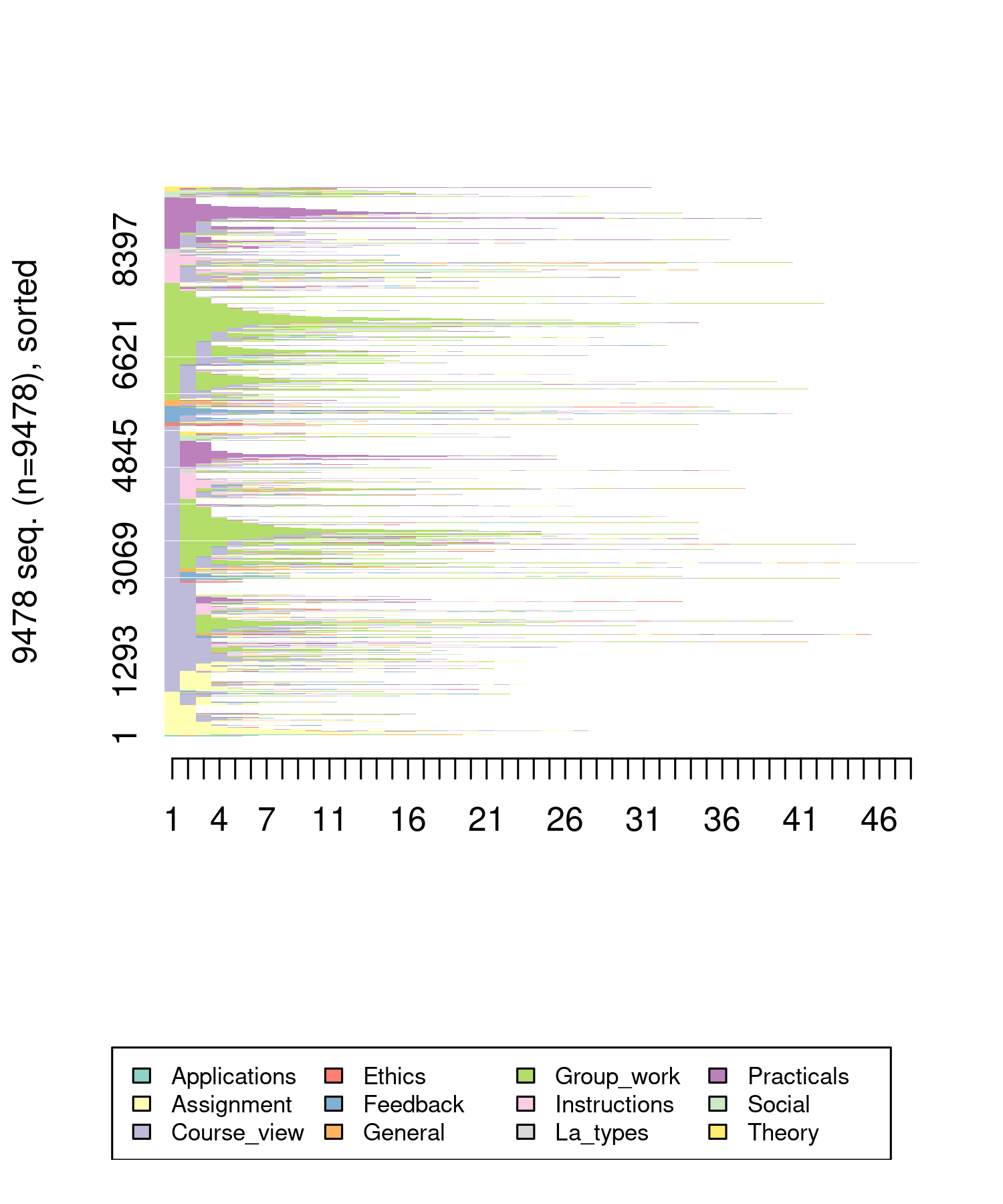}

}

\caption{\label{fig-process-index}Sequence index plot for the learning
management system logs.}

\end{figure}

Before proceeding further, it is advisable to visualise the sequences.
Figure~\ref{fig-process-index} shows the sequence index plot, sorted
according to the first states. The data are much larger than the
collaboration roles and achievement sequences analysed previously; there
are 9478 observations with an alphabet of 12 states. Unlike in the
previous example, the sequence lengths vary considerably. Due to this,
shorter sequences contain missing values to fill the empty cells in the
data frame. However, there are no internal gaps. When creating the
sequence object with the \texttt{seqdef} function, \texttt{TraMineR}
allows for distinguishing between real missing values (\texttt{NA},
where the true state is unknown) and technical missing values (void)
used to pad the sequences to equal lengths. The \texttt{seqHMM} package
is able to account for both types of missing values and treats them
slightly differently, for example when calculating the most probable
paths of hidden states.

\begin{Shaded}
\begin{Highlighting}[]
\FunctionTok{seqplot}\NormalTok{(seq\_data\_all, }\AttributeTok{type =} \StringTok{"I"}\NormalTok{, }\AttributeTok{ncol =} \DecValTok{4}\NormalTok{, }\AttributeTok{sortv =} \StringTok{"from.start"}\NormalTok{,}
        \AttributeTok{legend.prop =} \FloatTok{0.2}\NormalTok{, }\AttributeTok{cex.legend =} \FloatTok{0.7}\NormalTok{, }\AttributeTok{border =} \ConstantTok{NA}\NormalTok{)}
\end{Highlighting}
\end{Shaded}

A simple transition analysis can be performed by estimating and plotting
the transition probabilities. This can be performed using the
\texttt{TraMineR} package. Yet, this simple approach has drawbacks and
it is advisable to estimate the MM and use their full power. The next
code estimates the transition probabilities of the full dataset and
visualize them using the function \texttt{seqtrate()} from
\texttt{TraMineR} package.

\begin{Shaded}
\begin{Highlighting}[]
\NormalTok{overalltransitions }\OtherTok{\textless{}{-}} \FunctionTok{seqtrate}\NormalTok{(seq\_data\_all)}
\end{Highlighting}
\end{Shaded}

\begingroup 
\fontsize{4pt}{5pt}\selectfont
\addtolength{\tabcolsep}{-3pt}
\tiny

\hypertarget{tbl-transition}{}
\begin{longtable}{lrrrrrrrrrrrr}
\caption{\label{tbl-transition}Transition probabilities }\tabularnewline

\toprule
From\textbackslash{}To & Applications & Assignment & Course\_view & Ethics & Feedback & General & Group\_work & Instructions & La\_types & Practicals & Social & Theory \\ 
\midrule
Applications & $0.46$ & $0.07$ & $0.13$ & $0.01$ & $0.01$ & $0.19$ & $0.05$ & $0.01$ & $0.01$ & $0.05$ & $0.00$ & $0.00$ \\ 
Assignment & $0.00$ & $0.70$ & $0.19$ & $0.00$ & $0.01$ & $0.02$ & $0.03$ & $0.02$ & $0.02$ & $0.02$ & $0.00$ & $0.00$ \\ 
Course\_view & $0.01$ & $0.07$ & $0.35$ & $0.01$ & $0.03$ & $0.03$ & $0.28$ & $0.10$ & $0.02$ & $0.08$ & $0.02$ & $0.01$ \\ 
Ethics & $0.01$ & $0.00$ & $0.12$ & $0.61$ & $0.01$ & $0.04$ & $0.10$ & $0.01$ & $0.03$ & $0.04$ & $0.01$ & $0.02$ \\ 
Feedback & $0.00$ & $0.02$ & $0.23$ & $0.00$ & $0.56$ & $0.00$ & $0.11$ & $0.04$ & $0.01$ & $0.02$ & $0.00$ & $0.00$ \\ 
General & $0.04$ & $0.05$ & $0.18$ & $0.01$ & $0.00$ & $0.49$ & $0.06$ & $0.06$ & $0.05$ & $0.03$ & $0.01$ & $0.02$ \\ 
Group\_work & $0.00$ & $0.01$ & $0.19$ & $0.00$ & $0.01$ & $0.01$ & $0.73$ & $0.02$ & $0.00$ & $0.01$ & $0.01$ & $0.00$ \\ 
Instructions & $0.00$ & $0.02$ & $0.33$ & $0.00$ & $0.03$ & $0.04$ & $0.12$ & $0.37$ & $0.02$ & $0.03$ & $0.04$ & $0.00$ \\ 
La\_types & $0.01$ & $0.06$ & $0.24$ & $0.01$ & $0.00$ & $0.10$ & $0.07$ & $0.05$ & $0.38$ & $0.03$ & $0.01$ & $0.03$ \\ 
Practicals & $0.00$ & $0.02$ & $0.17$ & $0.00$ & $0.01$ & $0.01$ & $0.03$ & $0.02$ & $0.01$ & $0.73$ & $0.00$ & $0.01$ \\ 
Social & $0.00$ & $0.01$ & $0.25$ & $0.00$ & $0.00$ & $0.01$ & $0.12$ & $0.11$ & $0.01$ & $0.02$ & $0.48$ & $0.00$ \\ 
Theory & $0.00$ & $0.02$ & $0.15$ & $0.03$ & $0.00$ & $0.02$ & $0.06$ & $0.01$ & $0.05$ & $0.05$ & $0.00$ & $0.60$ \\ 
\bottomrule
\end{longtable}

\normalsize
\endgroup

\begin{figure}[H]

{\centering \includegraphics{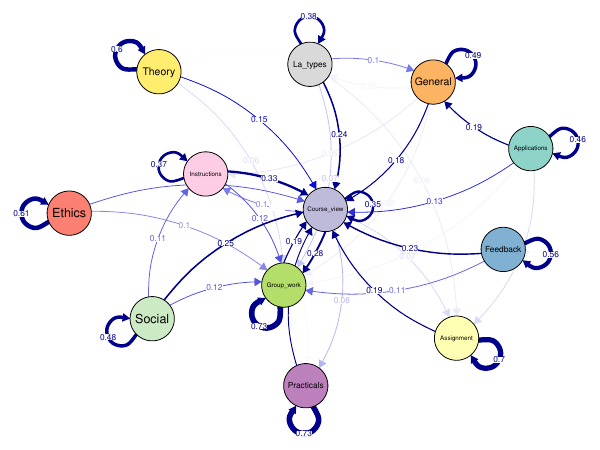}

}

\caption{\label{fig-overallplot1}Process map for the overall process.}

\end{figure}

As we mentioned earlier, we will use a novel plotting technique that is
more suitable for large process models. Below, we plot the transition
probabilities with the \texttt{qgraph()} function from the
\texttt{qgraph} package (Figure~\ref{fig-overallplot1}). We use some
arguments to improve the process model visualization. First, we use the
argument \texttt{cut\ =\ 0.15} to show the edges with probabilities
below 0.15 in lower thickness and color intensity. This \emph{cut} makes
the graph easier to read and less crowded, and gives emphasis to the
edges which matter. The argument \texttt{minimum\ =\ 0.05} hides small
edges below the probability threshold of 0.05. We use
\texttt{edge.labels\ =\ TRUE} to show the transition probabilities as
edge labels. The argument \texttt{color} gets the colour palette from
the sequence with the function \texttt{cpal()} and the argument
\texttt{curveAll\ =\ TRUE} ensures the graph shows curved edges. The
\texttt{"colorblind"} theme makes sure that the colours can be seen by
everyone regardless of colour vision abilities. Lastly, the \texttt{mar}
argument sets the margin of the figure to make all graphical aspects fit
within the figure area.

\begin{Shaded}
\begin{Highlighting}[]
\NormalTok{Labelx }\OtherTok{\textless{}{-}} \FunctionTok{alphabet}\NormalTok{(seq\_data\_all) }\CommentTok{\# get the labels to use them as nodes names.}
\NormalTok{transitionsplot }\OtherTok{\textless{}{-}} \FunctionTok{qgraph}\NormalTok{(overalltransitions, }\AttributeTok{cut =} \FloatTok{0.15}\NormalTok{, }\AttributeTok{minimum =} \FloatTok{0.05}\NormalTok{, }
                        \AttributeTok{labels =}\NormalTok{ Labelx, }\AttributeTok{edge.labels =} \ConstantTok{TRUE}\NormalTok{, }\AttributeTok{edge.label.cex =} \FloatTok{0.65}\NormalTok{, }
                        \AttributeTok{color =} \FunctionTok{cpal}\NormalTok{(seq\_data\_all), }\AttributeTok{curveAll =} \ConstantTok{TRUE}\NormalTok{, }
                        \AttributeTok{theme =} \StringTok{"colorblind"}\NormalTok{, }\AttributeTok{mar =} \FunctionTok{c}\NormalTok{(}\DecValTok{4}\NormalTok{, }\DecValTok{3}\NormalTok{, }\DecValTok{4}\NormalTok{, }\DecValTok{3}\NormalTok{))}
\end{Highlighting}
\end{Shaded}

The \texttt{seqtrate()} function only computes the transition
probabilities but does not compute the initial probabilities. While it
is not difficult to calculate the proportions of starting in each state,
we can also estimate a simple Markov model which does the same with a
short command. We do so using the \texttt{build\_mm()} function as per
Section \ref{markov}, recalling that the \texttt{build\_mm()} function
is distinct from \texttt{build\_hmm()}, \texttt{build\_mmm()}, and
\texttt{build\_mhmm()} in that it is the only build function that
automatically estimates the parameters of the model.

\begin{figure}[H]

{\centering \includegraphics{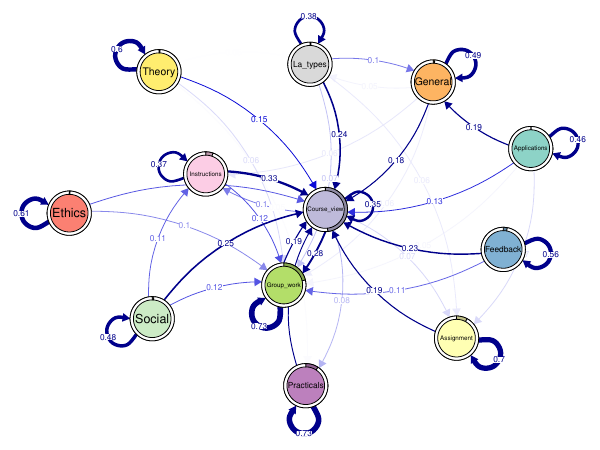}

}

\caption{\label{fig-overallplot2}Process map for the overall process
with initial probabilities.}

\end{figure}

The plotting now includes an extra option called
\texttt{pie\ =\ overallmodel\$initial\_probs} which tells
\texttt{qgraph} to use the initial probabilities from the fitted MM as
the sizes of the pie charts in the borders of the nodes in
Figure~\ref{fig-overallplot2}. For instance, the pie around \emph{Course
view} is around half of the circle corresponding to 0.48 initial
probability to start from \emph{Course view}. Please also note that the
graph is otherwise equal to the one generated via \texttt{seqtrate()}
apart from these initial probabilities.

\begin{Shaded}
\begin{Highlighting}[]
\NormalTok{overallmodel }\OtherTok{\textless{}{-}} \FunctionTok{build\_mm}\NormalTok{(seq\_data\_all)}

\NormalTok{overallplot }\OtherTok{\textless{}{-}} \FunctionTok{qgraph}\NormalTok{(overalltransitions, }
                      \AttributeTok{cut =} \FloatTok{0.15}\NormalTok{, }
                      \AttributeTok{minimum =} \FloatTok{0.05}\NormalTok{, }
                      \AttributeTok{labels =}\NormalTok{ Labelx, }
                      \AttributeTok{mar =} \FunctionTok{c}\NormalTok{(}\DecValTok{4}\NormalTok{, }\DecValTok{3}\NormalTok{, }\DecValTok{4}\NormalTok{, }\DecValTok{3}\NormalTok{), }
                      \AttributeTok{edge.labels =} \ConstantTok{TRUE}\NormalTok{, }
                      \AttributeTok{edge.label.cex =} \FloatTok{0.65}\NormalTok{, }
                      \AttributeTok{color =} \FunctionTok{cpal}\NormalTok{(seq\_data\_all), }
                      \AttributeTok{curveAll =} \ConstantTok{TRUE}\NormalTok{, }
                      \AttributeTok{theme =} \StringTok{"colorblind"}\NormalTok{, }
                      \AttributeTok{pie =}\NormalTok{ overallmodel}\SpecialCharTok{$}\NormalTok{initial\_probs)}
\end{Highlighting}
\end{Shaded}

Having plotted the transitions of the full dataset, we can now look for
transition patterns, that is typical transition patterns (i.e.,
clusters) that are repeated within the data. The procedure is the same
as before. In the next example, we use the function
\texttt{build\_mmm()} to build the model with four clusters as a
demonstration. Ideally, researchers need to estimate several models and
choose the best model based on model selection criteria (such as BIC)
values as well as interpretability.

The steps involved in fitting the model are as before; we make use of
the function \texttt{fit\_model()} to estimate the model. The results of
the running the code will be an MM for each cluster (with distinct
initial and transition probabilities). Given the number of sequences in
the dataset, their length, and the number of states, the computational
burden is larger than for previous applications in this chapter. For
illustrative purposes, instead of repeated EM runs with random starting
values, we use single EM run followed by global optimisation, using the
argument \texttt{global\_step\ =\ TRUE}. One benefit of this global (and
local) step in \texttt{fit\_model} over the EM algorithm is the
flexibility to define a maximum runtime (in seconds) for the
optimization process (argument \texttt{maxtime} in
\texttt{control\_global}). This can be valuable for larger problems with
predefined runtime (e.g., in a shared computer cluster). Note, however,
that relying on the runtime can lead to non-reproducible results even
with fixed seed if the optimisation terminates due to the time limit.
Finally, we run additional local optimisation step using the results of
the global optimisation, for more accurate results. The last argument
\texttt{threads\ =\ 16} instructs to use parallel computing to enable
faster fitting (please, customise according to the number of cores in
your computer). As for the starting values, we use the the transition
probabilities computed from the full data for all clusters, and random
values for the initial probabilities.

While in theory many of global optimisation algorithms should eventually
find the global optimum, in practice there are no guarantees that it is
found in limited time. Thus, as earlier, in practice it is advisable to
try different global/local optimisation algorithms and/or EM algorithm
with different initial values to make it more likely that the global
optimum is found (see {[}4{]} for further discussion).

\begin{Shaded}
\begin{Highlighting}[]
\FunctionTok{set.seed}\NormalTok{(}\DecValTok{1}\NormalTok{)}
\NormalTok{trans\_probs }\OtherTok{\textless{}{-}} \FunctionTok{simulate\_transition\_probs}\NormalTok{(}\DecValTok{12}\NormalTok{, }\DecValTok{4}\NormalTok{, }\AttributeTok{diag\_c =} \DecValTok{5}\NormalTok{)}
\NormalTok{init\_probs }\OtherTok{\textless{}{-}} \FunctionTok{as.numeric}\NormalTok{(}\FunctionTok{prop.table}\NormalTok{(}\FunctionTok{table}\NormalTok{(seq\_data\_all[,}\DecValTok{1}\NormalTok{])[}\DecValTok{1}\SpecialCharTok{:}\DecValTok{12}\NormalTok{]))}
\NormalTok{init\_probs }\OtherTok{\textless{}{-}} \FunctionTok{replicate}\NormalTok{(}\DecValTok{4}\NormalTok{, init\_probs, }\AttributeTok{simplify =} \ConstantTok{FALSE}\NormalTok{)}
\NormalTok{builtseqLMS }\OtherTok{\textless{}{-}} \FunctionTok{build\_mmm}\NormalTok{(seq\_data\_all,}
                         \AttributeTok{transition\_probs =}\NormalTok{ trans\_probs,}
                         \AttributeTok{initial\_probs =}\NormalTok{ init\_probs)}
\NormalTok{fitLMS }\OtherTok{\textless{}{-}} \FunctionTok{fit\_model}\NormalTok{(builtseqLMS, }
                    \AttributeTok{global\_step =} \ConstantTok{TRUE}\NormalTok{,}
                    \AttributeTok{control\_global =} \FunctionTok{list}\NormalTok{(}
                      \AttributeTok{maxtime =} \DecValTok{3600}\NormalTok{, }
                      \AttributeTok{maxeval =} \FloatTok{1e5}\NormalTok{,}
                      \AttributeTok{algorithm =} \StringTok{"NLOPT\_GD\_STOGO\_RAND"}\NormalTok{),}
                    \AttributeTok{local\_step =} \ConstantTok{TRUE}\NormalTok{,}
                    \AttributeTok{threads =} \DecValTok{16}\NormalTok{)}
\NormalTok{fitLMS}\SpecialCharTok{$}\NormalTok{global\_results}\SpecialCharTok{$}\NormalTok{message}
\NormalTok{fitLMS}\SpecialCharTok{$}\NormalTok{logLik}
\end{Highlighting}
\end{Shaded}

\begin{verbatim}
[1] "NLOPT_SUCCESS: Generic success return value."
\end{verbatim}

\begin{verbatim}
[1] -114491.2
\end{verbatim}

Before plotting the clusters, let us do some cleanups. First, we get the
transition probabilities of each cluster and assign them to a variable.
In that way, it is easier to manipulate and work with. In the same way,
we can extract the initial probabilities for each cluster.

\begin{Shaded}
\begin{Highlighting}[]
\CommentTok{\#extract transition probabilities of each cluster}
\NormalTok{Clustertp1 }\OtherTok{\textless{}{-}}\NormalTok{ fitLMS}\SpecialCharTok{$}\NormalTok{model}\SpecialCharTok{$}\NormalTok{transition\_probs}\SpecialCharTok{$}\StringTok{\textasciigrave{}}\AttributeTok{Cluster 1}\StringTok{\textasciigrave{}}
\NormalTok{Clustertp2 }\OtherTok{\textless{}{-}}\NormalTok{ fitLMS}\SpecialCharTok{$}\NormalTok{model}\SpecialCharTok{$}\NormalTok{transition\_probs}\SpecialCharTok{$}\StringTok{\textasciigrave{}}\AttributeTok{Cluster 2}\StringTok{\textasciigrave{}}
\NormalTok{Clustertp3 }\OtherTok{\textless{}{-}}\NormalTok{ fitLMS}\SpecialCharTok{$}\NormalTok{model}\SpecialCharTok{$}\NormalTok{transition\_probs}\SpecialCharTok{$}\StringTok{\textasciigrave{}}\AttributeTok{Cluster 3}\StringTok{\textasciigrave{}}
\NormalTok{Clustertp4 }\OtherTok{\textless{}{-}}\NormalTok{ fitLMS}\SpecialCharTok{$}\NormalTok{model}\SpecialCharTok{$}\NormalTok{transition\_probs}\SpecialCharTok{$}\StringTok{\textasciigrave{}}\AttributeTok{Cluster 4}\StringTok{\textasciigrave{}}

\CommentTok{\#extract initial probabilities of each cluster}
\NormalTok{Clusterinitp1 }\OtherTok{\textless{}{-}}\NormalTok{ fitLMS}\SpecialCharTok{$}\NormalTok{model}\SpecialCharTok{$}\NormalTok{initial\_probs}\SpecialCharTok{$}\StringTok{\textasciigrave{}}\AttributeTok{Cluster 1}\StringTok{\textasciigrave{}}
\NormalTok{Clusterinitp2 }\OtherTok{\textless{}{-}}\NormalTok{ fitLMS}\SpecialCharTok{$}\NormalTok{model}\SpecialCharTok{$}\NormalTok{initial\_probs}\SpecialCharTok{$}\StringTok{\textasciigrave{}}\AttributeTok{Cluster 2}\StringTok{\textasciigrave{}}
\NormalTok{Clusterinitp3 }\OtherTok{\textless{}{-}}\NormalTok{ fitLMS}\SpecialCharTok{$}\NormalTok{model}\SpecialCharTok{$}\NormalTok{initial\_probs}\SpecialCharTok{$}\StringTok{\textasciigrave{}}\AttributeTok{Cluster 3}\StringTok{\textasciigrave{}}
\NormalTok{Clusterinitp4 }\OtherTok{\textless{}{-}}\NormalTok{ fitLMS}\SpecialCharTok{$}\NormalTok{model}\SpecialCharTok{$}\NormalTok{initial\_probs}\SpecialCharTok{$}\StringTok{\textasciigrave{}}\AttributeTok{Cluster 4}\StringTok{\textasciigrave{}}
\end{Highlighting}
\end{Shaded}

\begin{figure}

{\centering \includegraphics{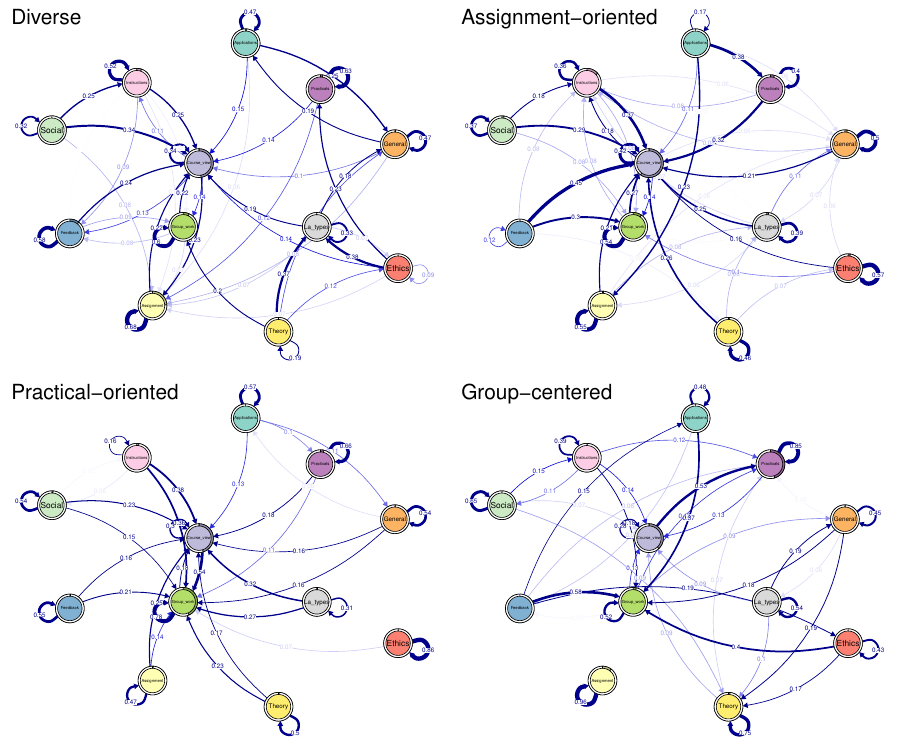}

}

\caption{\label{fig-plotclusters}Process maps for each cluster.}

\end{figure}

Plotting the process maps can be performed in the same way we did
before. However, if we need to compare clusters, it is best if we use a
unified layout. An average layout can be computed with the function
\texttt{averageLayout()} which takes the transition probabilities of the
four clusters as input and creates --- as the name implies--- an
averaged layout. Another option is to use the same layout of the
\texttt{overallplot} in the previous example. This can be obtained from
the plot object \texttt{overallplot\$layout}. This can be helpful if you
would like to plot the four plots corresponding to each cluster with the
same layout as the overall plot (see Figure~\ref{fig-plotclusters}).

\begin{Shaded}
\begin{Highlighting}[]
\NormalTok{Labelx }\OtherTok{\textless{}{-}} \FunctionTok{colnames}\NormalTok{(Clustertp1) }\CommentTok{\# we need to get the labels}

\NormalTok{Averagelayout }\OtherTok{\textless{}{-}} \FunctionTok{averageLayout}\NormalTok{(}\FunctionTok{list}\NormalTok{(Clustertp1, Clustertp2, Clustertp3, Clustertp4))}
\NormalTok{Overalllayout }\OtherTok{\textless{}{-}}\NormalTok{ overallplot}\SpecialCharTok{$}\NormalTok{layout }\CommentTok{\# You can also try with this layout from the previous plot}

\FunctionTok{qgraph}\NormalTok{(Clustertp1, }\AttributeTok{cut =} \FloatTok{0.15}\NormalTok{, }\AttributeTok{minimum =} \FloatTok{0.05}\NormalTok{ , }\AttributeTok{labels =}\NormalTok{ Labelx,}
       \AttributeTok{edge.labels =} \ConstantTok{TRUE}\NormalTok{, }\AttributeTok{edge.label.cex =} \FloatTok{0.65}\NormalTok{, }\AttributeTok{color =} \FunctionTok{cpal}\NormalTok{(seq\_data\_all), }
       \AttributeTok{layout =}\NormalTok{ Averagelayout, }\AttributeTok{pie =}\NormalTok{ Clusterinitp1, }\AttributeTok{curveAll =} \ConstantTok{TRUE}\NormalTok{, }
       \AttributeTok{theme =} \StringTok{"colorblind"}\NormalTok{, }\AttributeTok{title =} \StringTok{"Diverse"}\NormalTok{)}

\FunctionTok{qgraph}\NormalTok{(Clustertp2, }\AttributeTok{cut =} \FloatTok{0.15}\NormalTok{, }\AttributeTok{minimum =} \FloatTok{0.05}\NormalTok{, }\AttributeTok{labels =}\NormalTok{ Labelx,}
       \AttributeTok{edge.labels =} \ConstantTok{TRUE}\NormalTok{, }\AttributeTok{edge.label.cex =} \FloatTok{0.65}\NormalTok{, }\AttributeTok{color =} \FunctionTok{cpal}\NormalTok{(seq\_data\_all),  }
       \AttributeTok{layout =}\NormalTok{ Averagelayout, }\AttributeTok{pie =}\NormalTok{ Clusterinitp2, }\AttributeTok{curveAll =} \ConstantTok{TRUE}\NormalTok{, }
       \AttributeTok{theme =} \StringTok{"colorblind"}\NormalTok{, }\AttributeTok{title =} \StringTok{"Assignment{-}oriented"}\NormalTok{)}

\FunctionTok{qgraph}\NormalTok{(Clustertp3, }\AttributeTok{cut =} \FloatTok{0.15}\NormalTok{, }\AttributeTok{minimum =} \FloatTok{0.05}\NormalTok{, }\AttributeTok{labels =}\NormalTok{ Labelx,}
       \AttributeTok{edge.labels =} \ConstantTok{TRUE}\NormalTok{, }\AttributeTok{edge.label.cex =} \FloatTok{0.65}\NormalTok{, }\AttributeTok{color =} \FunctionTok{cpal}\NormalTok{(seq\_data\_all),  }
       \AttributeTok{layout =}\NormalTok{ Averagelayout, }\AttributeTok{pie =}\NormalTok{ Clusterinitp3, }\AttributeTok{curveAll =} \ConstantTok{TRUE}\NormalTok{, }
       \AttributeTok{theme =} \StringTok{"colorblind"}\NormalTok{, }\AttributeTok{title =} \StringTok{"Practical{-}oriented"}\NormalTok{)}

\FunctionTok{qgraph}\NormalTok{(Clustertp4, }\AttributeTok{cut =} \FloatTok{0.15}\NormalTok{, }\AttributeTok{minimum =} \FloatTok{0.05}\NormalTok{ , }\AttributeTok{labels =}\NormalTok{ Labelx, }
       \AttributeTok{edge.labels =} \ConstantTok{TRUE}\NormalTok{, }\AttributeTok{edge.label.cex =} \FloatTok{0.65}\NormalTok{, }\AttributeTok{color =} \FunctionTok{cpal}\NormalTok{(seq\_data\_all),  }
       \AttributeTok{layout =}\NormalTok{ Averagelayout, }\AttributeTok{pie =}\NormalTok{ Clusterinitp4, }\AttributeTok{curveAll =} \ConstantTok{TRUE}\NormalTok{, }
       \AttributeTok{theme =} \StringTok{"colorblind"}\NormalTok{, }\AttributeTok{title =} \StringTok{"Group{-}centered"}\NormalTok{)}
\end{Highlighting}
\end{Shaded}

Oftentimes, the researcher is interested in comparing two pre-defined
fixed groups, e.g., high achievers and low achievers, rather than
between the computed clusters. In the next example we will compare high
to low achievers based on their achievement levels. First, we have to
create a separate sequence object for each group. We do this by
filtering but you can do it in other ways. For instance, you can create
two sequences from scratch for each group. The next is to build the MMs
separately for each group.

\begin{Shaded}
\begin{Highlighting}[]
\NormalTok{seq\_high }\OtherTok{\textless{}{-}}\NormalTok{ seq\_data\_all[seq\_data}\SpecialCharTok{$}\NormalTok{Achievementlevel4 }\SpecialCharTok{\textless{}=} \DecValTok{2}\NormalTok{,]}
\NormalTok{seq\_low }\OtherTok{\textless{}{-}}\NormalTok{  seq\_data\_all[seq\_data}\SpecialCharTok{$}\NormalTok{Achievementlevel4 }\SpecialCharTok{\textgreater{}} \DecValTok{2}\NormalTok{,]}

\NormalTok{high\_mm }\OtherTok{\textless{}{-}} \FunctionTok{build\_mm}\NormalTok{(seq\_high)}
\NormalTok{low\_mm }\OtherTok{\textless{}{-}} \FunctionTok{build\_mm}\NormalTok{(seq\_low)}
\end{Highlighting}
\end{Shaded}

\begin{figure}[H]

{\centering \includegraphics{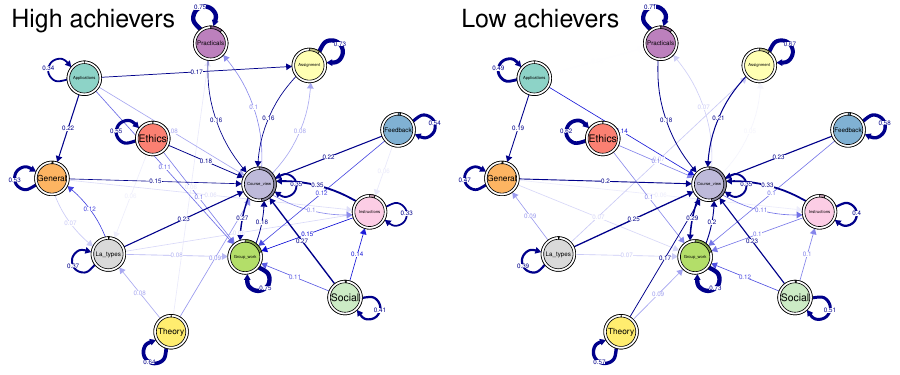}

}

\caption{\label{fig-hilow}Process maps for high achievers and low
achievers using average layout.}

\end{figure}

Before plotting the groups, let us do some cleaning, like we did before.
First, we get the transition and initial probabilities of each group. We
also compute an average layout. Please note that you can use the layout
from the previous examples if you are comparing the models against each
other and you need a unified framework. The plotting is the same as
before (see Figure~\ref{fig-hilow}).

\begin{Shaded}
\begin{Highlighting}[]
\FunctionTok{par}\NormalTok{(}\AttributeTok{mfrow=}\FunctionTok{c}\NormalTok{(}\DecValTok{1}\NormalTok{, }\DecValTok{2}\NormalTok{))}

\CommentTok{\#extract transition probabilities of each cluster}
\NormalTok{Highprobs }\OtherTok{\textless{}{-}}\NormalTok{ high\_mm}\SpecialCharTok{$}\NormalTok{transition\_probs}
\NormalTok{Lowprobs }\OtherTok{\textless{}{-}}\NormalTok{ low\_mm}\SpecialCharTok{$}\NormalTok{transition\_probs}

\CommentTok{\#extract initial probabilities of each cluster}
\NormalTok{Highinit }\OtherTok{\textless{}{-}}\NormalTok{ high\_mm}\SpecialCharTok{$}\NormalTok{initial\_probs}
\NormalTok{Lowinit }\OtherTok{\textless{}{-}}\NormalTok{ high\_mm}\SpecialCharTok{$}\NormalTok{initial\_probs}

\NormalTok{Averagelayout }\OtherTok{\textless{}{-}} \FunctionTok{averageLayout}\NormalTok{(}\FunctionTok{list}\NormalTok{(Highprobs, Lowprobs))}

\NormalTok{Highplot }\OtherTok{\textless{}{-}} \FunctionTok{qgraph}\NormalTok{(Highprobs, }\AttributeTok{cut =} \FloatTok{0.15}\NormalTok{, }\AttributeTok{minimum =} \FloatTok{0.05}\NormalTok{, }\AttributeTok{labels =}\NormalTok{ Labelx,}
                   \AttributeTok{edge.labels =} \ConstantTok{TRUE}\NormalTok{, }\AttributeTok{edge.label.cex =} \FloatTok{0.65}\NormalTok{, }
                   \AttributeTok{color =} \FunctionTok{cpal}\NormalTok{(seq\_data\_all), }\AttributeTok{layout =}\NormalTok{ Averagelayout, }
                   \AttributeTok{pie =}\NormalTok{ Highinit, }\AttributeTok{theme =} \StringTok{"colorblind"}\NormalTok{, }\AttributeTok{title =} \StringTok{"High achievers"}\NormalTok{)}

\NormalTok{Lowplot }\OtherTok{\textless{}{-}}  \FunctionTok{qgraph}\NormalTok{(Lowprobs, }\AttributeTok{cut=}\FloatTok{0.15}\NormalTok{, }\AttributeTok{minimum =} \FloatTok{0.05}\NormalTok{, }\AttributeTok{labels =}\NormalTok{ Labelx,}
                   \AttributeTok{edge.labels =} \ConstantTok{TRUE}\NormalTok{, }\AttributeTok{edge.label.cex =} \FloatTok{0.65}\NormalTok{, }
                   \AttributeTok{color =} \FunctionTok{cpal}\NormalTok{(seq\_data\_all), }\AttributeTok{layout =}\NormalTok{ Averagelayout, }
                   \AttributeTok{pie =}\NormalTok{ Lowinit, }\AttributeTok{theme =} \StringTok{"colorblind"}\NormalTok{, }\AttributeTok{title =} \StringTok{"Low achievers"}\NormalTok{)}
\end{Highlighting}
\end{Shaded}

\begin{figure}[H]

{\centering \includegraphics{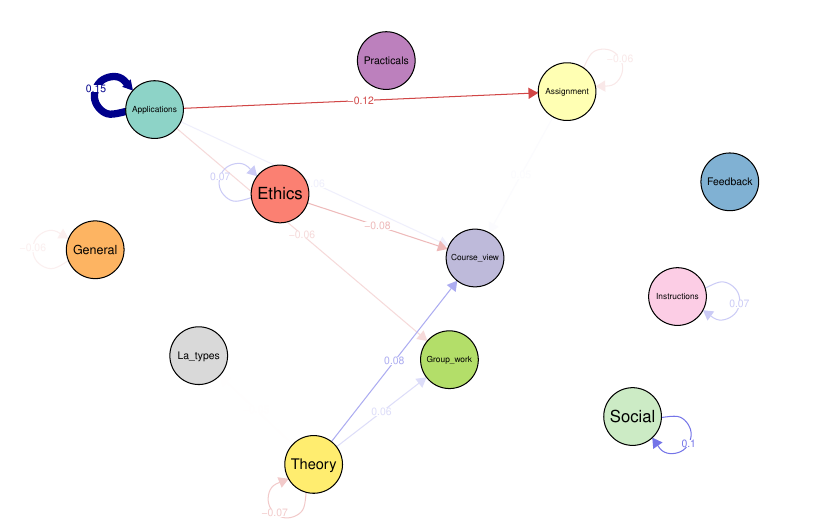}

}

\caption{\label{fig-diff}Difference between process maps of high
achievers and low achievers using average layout.}

\end{figure}

We can also plot the difference plot (see Figure~\ref{fig-diff}); that
is, what the low achievers do less than high achievers. In this case,
red edges are negative (events they do less) and blue edges are positive
(events that they do more than high achievers). As you can see, the
differences are not that huge. In fact, much of the literature comparing
high and low achievers uses higher thresholds e.g., top 25\% to bottom
25\% or even top 10\% to bottom 10\%.

\begin{Shaded}
\begin{Highlighting}[]
\NormalTok{diffplot }\OtherTok{\textless{}{-}} \FunctionTok{qgraph}\NormalTok{(Lowprobs }\SpecialCharTok{{-}}\NormalTok{ Highprobs, }\AttributeTok{cut =} \FloatTok{0.15}\NormalTok{, }\AttributeTok{minimum =} \FloatTok{0.05}\NormalTok{, }\AttributeTok{labels =}\NormalTok{ Labelx,}
                   \AttributeTok{edge.labels =} \ConstantTok{TRUE}\NormalTok{, }\AttributeTok{edge.label.cex =} \FloatTok{0.65}\NormalTok{, }\AttributeTok{layout =}\NormalTok{ Averagelayout, }
                   \AttributeTok{color =} \FunctionTok{cpal}\NormalTok{(seq\_data\_all), }\AttributeTok{theme =} \StringTok{"colorblind"}\NormalTok{)}
\end{Highlighting}
\end{Shaded}

\hypertarget{conclusions-further-readings}{%
\section{Conclusions \& further
readings}\label{conclusions-further-readings}}

Markovian models provide a flexible model-based approach for analysing
complex sequence data. MMs and HMMs have proven useful in many
application areas such as biology and speech recognition, and can be a
valuable tool in analysing data in educational settings as well. Their
mixture variants allow for the representation of complex systems by
combining multiple MMs or HMMs, each capturing different aspects of the
underlying processes, allowing probabilistic clustering, information
compression (e.g.~visualisation of multicategory data from multiple
domains), and detection of latent features of sequence data (e.g,
extraction of different learning strategies). The ability to incorporate
covariates in the case of MMMs and MHMMs makes those models even more
powerful, and generally MMs and MMMs represent useful tools in the field
of process mining also.

The \texttt{seqHMM} package used in the examples supports time-constant
covariates for predicting cluster memberships for each individual. In
theory, covariates could be used to define transition or emission
probabilities as well, leading to subject-specific and possibly
time-varying transition and emission probabilities (in the case of
time-varying covariates). However, at the time of writing this chapter,
these are not supported in \texttt{seqHMM} (this may change in the
future). In R, there are at least two other, potentially useful
packages: for MMs, the \texttt{dynamite} {[}38{]} package supports
covariates on the transition probabilities with potentially time-varying
effects, whereas \texttt{LMest} {[}39{]} supports MMs and HMMs with
covariates, and restricted variants of the MHMM where only the initial
and transition matrices vary between clusters. Going beyond the R
software, some commercial software also offers tools to analyse
Markovian models, including latentGold {[}40{]} and Mplus {[}11{]}.

The conditional independence assumption of observations given the latent
states in HMMs can sometimes be unrealistic. In these settings, the
so-called double chain MMs can be used {[}41{]}. There the current
observation is allowed to depend on both the current state and the
previous observation. Some restricted variants of such models are
implemented in the \texttt{march} package in R {[}42{]}. Finally,
variable order MMs extend basic MMs by allowing the order of the MM to
vary in time. A \texttt{TraMineR}-compatible implementation of
variable-order models can be found in the \texttt{PST} package {[}43{]}.

We encourage readers to read more about how to interpret the results in
the original study where the data for this chapter was drawn from
{[}36{]}. We also encourage readers to learn more about Markovian models
in the context of multi-channel sequence analysis in Chapter 13 {[}8{]}.

\hypertarget{references}{%
\section{References}\label{references}}

\hypertarget{refs}{}
\begin{CSLReferences}{0}{0}
\leavevmode\vadjust pre{\hypertarget{ref-Saqr2024-tv}{}}%
\CSLLeftMargin{1. }%
\CSLRightInline{Saqr M, López-Pernas S, Helske S, Durand M, Murphy K,
Studer M, Ritschard G (2024) Sequence analysis: Basic principles,
technique, and tutorial. In: Saqr M, López-Pernas S (eds) Learning
analytics methods and tutorials: A practical guide using {R}. Springer}

\leavevmode\vadjust pre{\hypertarget{ref-LopezPernas2024}{}}%
\CSLLeftMargin{2. }%
\CSLRightInline{López-Pernas S, Saqr M (2024) Modelling the dynamics of
longitudinal processes in education. A tutorial with {R} for the
{VaSSTra} method. In: Saqr M, López-Pernas S (eds) Learning analytics
methods and tutorials: A practical guide using {R}. Springer, pp
in--press}

\leavevmode\vadjust pre{\hypertarget{ref-Liao2022}{}}%
\CSLLeftMargin{3. }%
\CSLRightInline{Liao TF, Bolano D, Brzinsky-Fay C, Cornwell B, Fasang
AE, Helske S, Piccarreta R, Raab M, Ritschard G, Struffolino E, Studer M
(2022) Sequence analysis: Its past, present, and future. Social Science
Research 107:102772.
https://doi.org/\href{https://doi.org/10.1016/j.ssresearch.2022.102772}{10.1016/j.ssresearch.2022.102772}}

\leavevmode\vadjust pre{\hypertarget{ref-Helske2019}{}}%
\CSLLeftMargin{4. }%
\CSLRightInline{Helske S, Helske J (2019) Mixture hidden {M}arkov models
for sequence data: The {seqHMM} package in {R}. Journal of Statistical
Software 88:
https://doi.org/\href{https://doi.org/10.18637/jss.v088.i03}{10.18637/jss.v088.i03}}

\leavevmode\vadjust pre{\hypertarget{ref-schwarz1978}{}}%
\CSLLeftMargin{5. }%
\CSLRightInline{Schwarz GE (1978) Estimating the dimension of a model.
The Annals of Statistics 6:461--464.
https://doi.org/\href{https://doi.org/10.1214/aos/1176344136}{10.1214/aos/1176344136}}

\leavevmode\vadjust pre{\hypertarget{ref-vandePol1990}{}}%
\CSLLeftMargin{6. }%
\CSLRightInline{Pol F van de, Langeheine R (1990) Mixed {M}arkov latent
class models. Sociological Methodology 20:213.
https://doi.org/\href{https://doi.org/10.2307/271087}{10.2307/271087}}

\leavevmode\vadjust pre{\hypertarget{ref-Vermunt}{}}%
\CSLLeftMargin{7. }%
\CSLRightInline{Vermunt JK, Tran B, Magidson J (2008) Latent class
models in longitudinal research. In: Menard S (ed) Handbook of
longitudinal research. Elsevier, Netherlands, pp 373--385}

\leavevmode\vadjust pre{\hypertarget{ref-Lopez-Pernas2024-kf}{}}%
\CSLLeftMargin{8. }%
\CSLRightInline{López-Pernas S, Saqr M, Murphy K (2024) Multi-channel
sequence analysis in educational research: An introduction and tutorial
with {R}. In: Saqr M, López-Pernas S (eds) Learning analytics methods
and tutorials: A practical guide using {R}. Springer, pp in--press}

\leavevmode\vadjust pre{\hypertarget{ref-Rabiner1989}{}}%
\CSLLeftMargin{9. }%
\CSLRightInline{Rabiner L (1989) A tutorial on hidden {M}arkov models
and selected applications in speech recognition. Proceedings of the IEEE
77:257--286.
https://doi.org/\href{https://doi.org/10.1109/5.18626}{10.1109/5.18626}}

\leavevmode\vadjust pre{\hypertarget{ref-Helske2018}{}}%
\CSLLeftMargin{10. }%
\CSLRightInline{Helske S, Helske J, Eerola M (2018)
\href{https://doi.org/10.1007/978-3-319-95420-2_11}{Combining sequence
analysis and hidden markov models in the analysis of complex life
sequence data}. In: Ritschard G, Studer M (eds) Sequence analysis and
related approaches: Innovative methods and applications. Springer
International Publishing, Cham, pp 185--200}

\leavevmode\vadjust pre{\hypertarget{ref-mplus}{}}%
\CSLLeftMargin{11. }%
\CSLRightInline{Muthén LK, O. MB (1998-\/-2017) {Mplus User's Guide},
Eighth edition. Muthén \& Muthén, Los Angeles, CA, U.S.A.}

\leavevmode\vadjust pre{\hypertarget{ref-Muthen}{}}%
\CSLLeftMargin{12. }%
\CSLRightInline{Muthén B, Muthén L
\href{https://www.statmodel.com/download/Mplus-A\%20General\%20Latent\%20Variable\%20Modeling\%20Program.pdf}{Mplus:
A general latent variable modeling program}}

\leavevmode\vadjust pre{\hypertarget{ref-Tormanen2022-ux}{}}%
\CSLLeftMargin{13. }%
\CSLRightInline{Törmänen, Järvenoja, Saqr, Malmberg, others (2022) A
person-centered approach to study students' socio-emotional interaction
profiles and regulation of collaborative learning. Frontiers in
Education}

\leavevmode\vadjust pre{\hypertarget{ref-Tormanen2023-gz}{}}%
\CSLLeftMargin{14. }%
\CSLRightInline{Törmänen T, Järvenoja H, Saqr M, Malmberg J, Järvelä S
(2023) Affective states and regulation of learning during
socio-emotional interactions in secondary school collaborative groups.
British Journal of Educational Psychology 93 Suppl 1:48--70.
https://doi.org/\href{https://doi.org/10.1111/bjep.12525}{10.1111/bjep.12525}}

\leavevmode\vadjust pre{\hypertarget{ref-Fincham2019-yz}{}}%
\CSLLeftMargin{15. }%
\CSLRightInline{Fincham E, Gašević D, Jovanović J, Pardo A (2019) From
study tactics to learning strategies: An analytical method for
extracting interpretable representations. IEEE Transactions on Learning
Technologies 12:59--72.
https://doi.org/\href{https://doi.org/10.1109/TLT.2018.2823317}{10.1109/TLT.2018.2823317}}

\leavevmode\vadjust pre{\hypertarget{ref-Roles}{}}%
\CSLLeftMargin{16. }%
\CSLRightInline{Saqr M, López-Pernas S (2022) How {CSCL} roles emerge,
persist, transition, and evolve over time: A four-year longitudinal
study. Computers \& Education 189:104581}

\leavevmode\vadjust pre{\hypertarget{ref-IHE}{}}%
\CSLLeftMargin{17. }%
\CSLRightInline{Saqr M, López-Pernas S, Jovanović J, Gašević D (2023)
Intense, turbulent, or wallowing in the mire: A longitudinal study of
cross-course online tactics, strategies, and trajectories. The Internet
and Higher Education 57:100902}

\leavevmode\vadjust pre{\hypertarget{ref-BOUGUETTAYA20152785}{}}%
\CSLLeftMargin{18. }%
\CSLRightInline{Bouguettaya A, Yu Q, Liu X, Zhou X, Song A (2015)
Efficient agglomerative hierarchical clustering. Expert Systems with
Applications 42:2785--2797.
https://doi.org/\url{https://doi.org/10.1016/j.eswa.2014.09.054}}

\leavevmode\vadjust pre{\hypertarget{ref-Gilpin13}{}}%
\CSLLeftMargin{19. }%
\CSLRightInline{Gilpin S, Qian B, Davidson I (2013)
\href{https://doi.org/10.1145/2505515.2505527}{Efficient hierarchical
clustering of large high dimensional datasets}. In: Proceedings of the
22nd ACM international conference on information \& knowledge
management. Association for Computing Machinery, New York, NY, USA, pp
1371--1380}

\leavevmode\vadjust pre{\hypertarget{ref-Bringing}{}}%
\CSLLeftMargin{20. }%
\CSLRightInline{López-Pernas S, Saqr M (2021) Bringing synchrony and
clarity to complex multi-channel data: A learning analytics study in
programming education. IEEE Access 9:}

\leavevmode\vadjust pre{\hypertarget{ref-Engagement}{}}%
\CSLLeftMargin{21. }%
\CSLRightInline{Saqr M, López-Pernas S (2021) The longitudinal
trajectories of online engagement over a full program. Computers \&
Education 175:104325}

\leavevmode\vadjust pre{\hypertarget{ref-Matcha2020-jp}{}}%
\CSLLeftMargin{22. }%
\CSLRightInline{Matcha W, Gašević D, Ahmad Uzir N, Jovanović J, Pardo A,
Lim L, Maldonado-Mahauad J, Gentili S, Pérez-Sanagustín M, Tsai Y-S
(2020) Analytics of learning strategies: Role of course design and
delivery modality. Journal of Learning Analytics 7:45--71.
https://doi.org/\href{https://doi.org/10.18608/jla.2020.72.3}{10.18608/jla.2020.72.3}}

\leavevmode\vadjust pre{\hypertarget{ref-Peeters2020-wa}{}}%
\CSLLeftMargin{23. }%
\CSLRightInline{Peeters W, Saqr M, Viberg O (2020) Applying learning
analytics to map students' self-regulated learning tactics in an
academic writing course. In: Proceedings of the 28th international
conference on computers in education. pp 245--254}

\leavevmode\vadjust pre{\hypertarget{ref-Lim2023-kg}{}}%
\CSLLeftMargin{24. }%
\CSLRightInline{Lim L, Bannert M, Graaf J van der, Singh S, Fan Y,
Surendrannair S, Rakovic M, Molenaar I, Moore J, Gašević D (2023)
Effects of real-time analytics-based personalized scaffolds on students'
self-regulated learning. Computers in Human Behavior 139:107547.
https://doi.org/\href{https://doi.org/10.1016/j.chb.2022.107547}{10.1016/j.chb.2022.107547}}

\leavevmode\vadjust pre{\hypertarget{ref-Saqr2023-he}{}}%
\CSLLeftMargin{25. }%
\CSLRightInline{Saqr M, López-Pernas S (2023) The temporal dynamics of
online problem-based learning: Why and when sequence matters.
International Journal of Computer-Supported Collaborative Learning
18:11--37.
https://doi.org/\href{https://doi.org/10.1007/s11412-023-09385-1}{10.1007/s11412-023-09385-1}}

\leavevmode\vadjust pre{\hypertarget{ref-Gatta2017-rg}{}}%
\CSLLeftMargin{26. }%
\CSLRightInline{Gatta R, Vallati M, Lenkowicz J, Rojas E, Damiani A,
Sacchi L, De Bari B, Dagliati A, Fernandez-Llatas C, Montesi M,
Marchetti A, Castellano M, Valentini V (2017)
\href{https://doi.org/10.1145/3148011.3154464}{Generating and comparing
knowledge graphs of medical processes using {pMineR}}. In: Proceedings
of the knowledge capture conference. ACM, New York, NY, USA}

\leavevmode\vadjust pre{\hypertarget{ref-Boroujeni2019-vf}{}}%
\CSLLeftMargin{27. }%
\CSLRightInline{Boroujeni MS, Dillenbourg P (2019) Discovery and
temporal analysis of {MOOC} study patterns. Journal of Learning
Analytics 6:16--33.
https://doi.org/\href{https://doi.org/10.18608/jla.2019.61.2}{10.18608/jla.2019.61.2}}

\leavevmode\vadjust pre{\hypertarget{ref-Andrade2017-we}{}}%
\CSLLeftMargin{28. }%
\CSLRightInline{Andrade A, Danish JA, Maltese AV (2017) A measurement
model of gestures in an embodied learning environment: Accounting for
temporal dependencies. Journal of Learning Analytics 4:18--46.
https://doi.org/\href{https://doi.org/10.18608/jla.2017.43.3}{10.18608/jla.2017.43.3}}

\leavevmode\vadjust pre{\hypertarget{ref-Kokoc2021-rc}{}}%
\CSLLeftMargin{29. }%
\CSLRightInline{Kokoç M, Akçapınar G, Hasnine MN (2021)
\href{https://www.jstor.org/stable/26977869}{Unfolding students' online
assignment submission behavioral patterns using temporal learning
analytics}. Educational Technology \& Society 24:223--235}

\leavevmode\vadjust pre{\hypertarget{ref-qgraph}{}}%
\CSLLeftMargin{30. }%
\CSLRightInline{Epskamp S, Cramer AOJ, Waldorp LJ, Schmittmann VD,
Borsboom D (2012) {qgraph: network visualizations of relationships in
psychometric data}. Journal of Statistical Software 48:}

\leavevmode\vadjust pre{\hypertarget{ref-rio}{}}%
\CSLLeftMargin{31. }%
\CSLRightInline{Chan C, Chan GC, Leeper TJ, Becker J (2021) {rio: a
Swiss-army knife for data file I/O}}

\leavevmode\vadjust pre{\hypertarget{ref-seqHMM}{}}%
\CSLLeftMargin{32. }%
\CSLRightInline{Helske J, Helske S (2023)
\href{https://cran.r-project.org/package=seqHMM}{{seqHMM}: Mixture
hidden {Markov} models for social sequence data and other multivariate,
multichannel categorical time series}}

\leavevmode\vadjust pre{\hypertarget{ref-tidyverse}{}}%
\CSLLeftMargin{33. }%
\CSLRightInline{Wickham H, Averick M, Bryan J, Chang W, McGowan LD,
François R, Grolemund G, Hayes A, Henry L, Hester J, Kuhn M, Pedersen
TL, Miller E, Bache SM, Müller K, Ooms J, Robinson D, Seidel DP, Spinu
V, Takahashi K, Vaughan D, Wilke C, Woo K, Yutani H (2019) Welcome to
the {tidyverse}. Journal of Open Source Software 4:1686.
https://doi.org/\href{https://doi.org/10.21105/joss.01686}{10.21105/joss.01686}}

\leavevmode\vadjust pre{\hypertarget{ref-Gabadinho2011}{}}%
\CSLLeftMargin{34. }%
\CSLRightInline{Gabadinho A, Ritschard G, Müller NS, Studer M (2011)
Analyzing and visualizing state sequences in {R} with {TraMineR}.
Journal of Statistical Software 40:
https://doi.org/\href{https://doi.org/10.18637/jss.v040.i04}{10.18637/jss.v040.i04}}

\leavevmode\vadjust pre{\hypertarget{ref-Satu}{}}%
\CSLLeftMargin{35. }%
\CSLRightInline{Saqr M, López-Pernas S, Helske S, Hrastinski S (2023)
The longitudinal association between engagement and achievement varies
by time, students' profiles, and achievement state: A full program
study. Computers \& Education 199:104787}

\leavevmode\vadjust pre{\hypertarget{ref-SAQR2022104581}{}}%
\CSLLeftMargin{36. }%
\CSLRightInline{Saqr M, López-Pernas S (2022) How {CSCL} roles emerge,
persist, transition, and evolve over time: A four-year longitudinal
study. Computers \& Education 189:104581.
https://doi.org/\url{https://doi.org/10.1016/j.compedu.2022.104581}}

\leavevmode\vadjust pre{\hypertarget{ref-Helske2023}{}}%
\CSLLeftMargin{37. }%
\CSLRightInline{Helske S, Keski-Säntti M, Kivelä J, Juutinen A, Kääriälä
A, Gissler M, Merikukka M, Lallukka T (2023) Predicting the stability of
early employment with its timing and childhood social and health-related
predictors: A mixture markov model approach. Longitudinal and Life
Course Studies 14:73--104}

\leavevmode\vadjust pre{\hypertarget{ref-dynamitepaper}{}}%
\CSLLeftMargin{38. }%
\CSLRightInline{Tikka S, Helske J (2023)
\href{https://doi.org/10.48550/ARXIV.2302.01607}{{dynamite}: An {R}
package for dynamic multivariate panel models}}

\leavevmode\vadjust pre{\hypertarget{ref-LMest}{}}%
\CSLLeftMargin{39. }%
\CSLRightInline{Bartolucci F, Pandolfi S, Pennoni F (2017) {LMest}: An
{R} package for latent {M}arkov models for longitudinal categorical
data. Journal of Statistical Software 81:1--38.
https://doi.org/\href{https://doi.org/10.18637/jss.v081.i04}{10.18637/jss.v081.i04}}

\leavevmode\vadjust pre{\hypertarget{ref-latentgold}{}}%
\CSLLeftMargin{40. }%
\CSLRightInline{Vermunt JK, Magidson J (2016) {Guide for Latent GOLD
5.1: Basic, Advanced, and Syntax}. Statistical Innovations Inc.,
Belmont, MA, U.S.A.}

\leavevmode\vadjust pre{\hypertarget{ref-Berchtold}{}}%
\CSLLeftMargin{41. }%
\CSLRightInline{Berchtold A (1999) The double chain {M}arkov model.
Communications in Statistics - Theory and Methods 28:2569--2589.
https://doi.org/\href{https://doi.org/10.1080/03610929908832439}{10.1080/03610929908832439}}

\leavevmode\vadjust pre{\hypertarget{ref-march}{}}%
\CSLLeftMargin{42. }%
\CSLRightInline{Maitre O, Emery K, Oliver Buschor with contributions
from, Berchtold A (2020)
\href{https://CRAN.R-project.org/package=march}{{march: Markov chains}}}

\leavevmode\vadjust pre{\hypertarget{ref-Gabadinho2016}{}}%
\CSLLeftMargin{43. }%
\CSLRightInline{Gabadinho A, Ritschard G (2016) Analyzing state
sequences with probabilistic suffix trees: The {PST} {R} package.
Journal of Statistical Software 72:1--39.
https://doi.org/\href{https://doi.org/10.18637/jss.v072.i03}{10.18637/jss.v072.i03}}

\end{CSLReferences}

\end{document}